\documentclass[a4paper,12pt]{article}
\pdfoutput=1
\usepackage{jheppub}
\usepackage{amsmath,amssymb,bm}
\usepackage{amsthm,mathrsfs}
\usepackage{graphicx}
\usepackage{hepunits}
\usepackage[svgnames]{xcolor}
\usepackage{multirow}
\usepackage{float}
\usepackage{textcomp}
\usepackage{makecell}
\usepackage{cleveref}

\allowdisplaybreaks

\title{The impact of data from future lepton colliders on light hadrons fragmentation functions}
\author[a,b]{Bin Zhou}
\author[a,b]{and Jun Gao}

\affiliation[a]{INPAC, Shanghai Key Laboratory for Particle Physics and Cosmology,
School of Physics and Astronomy, Shanghai Jiao Tong University, Shanghai 200240, China}
\affiliation[b]{Key Laboratory for Particle Astrophysics and Cosmology, Shanghai 200240, China}

\emailAdd{zb0429@sjtu.edu.cn}
\emailAdd{jung49@sjtu.edu.cn}

\abstract{%
In this work, we study the constraining power of future lepton colliders on fragmentation functions (FFs) to light charged hadrons from quarks and gluon in the framework of QCD collinear factorization.
We perform analyses of FFs at NLO by including a wide range of pseudo--data from future lepton colliders, such as measurements on hadron multiplicities in the production of two jets and $W$ boson pairs, at various center of mass energies, and from hadronic decays of the Higgs boson, including both to heavy quarks and to gluons.
The high luminosity and high energies of future lepton colliders allow for quark flavor separations and ensure a precise determination of FFs based solely on data from electron-positron collisions.
We find that either the CEPC, FCC-$ee$ or ILC can significantly reduce the uncertainties of FFs in a wide kinematic range, compared to the NPC23 set obtained with a global analysis to current world data.
We also discuss the impact of higher-order QCD corrections, and the potential constraints from measurements of three-jet production.
Furthermore, we describe an update of the FMNLO program allowing for calculating hadron production cross sections at next-to-next-to-leading order in QCD, which is used in this study.}

\begin{document}

\maketitle

\clearpage

\section{Introduction}
A comprehensive grasp of the hadronization of quarks and gluons into hadrons is a fundamental element of theoretical predictions for specific high-energy processes with identified hadrons, including single-inclusive hadron production in electron-positron annihilation (SIA), semi-inclusive deep inelastic lepton-nucleon scattering (SIDIS) and single-inclusive hadron production in proton-proton (pp) collisions. This is quantified by fragmentation functions (FFs), which, in the simplest picture, describe the probability distribution on the fraction of the initial parton momentum carried by the identified hadron~\cite{Collins:1989gx,Collins:1998rz,Metz:2016swz}.

As the FFs are related to the nonperturbative aspect of QCD, they cannot be calculated from the first principle of QCD at present and are in general extracted from fits to a variety of experimental data. Many studies have already been conducted on a comprehensive fit involving various data samples to extract FFs.
The representative efforts in this field can be found in the works of DSS~\cite{deFlorian:2007ekg}, AKK~\cite{Albino:2008fy}, NNFF~\cite{Bertone:2018ecm}, MAPFF~\cite{Khalek:2021gxf}, JAM~\cite{Moffat:2021dji}, and NPC23~\cite{Gao:2024nkz,Gao:2024dbv}. They use all data sets from SIA, SIDIS, or pp collisions. In contrast, the HKNS analysis only includes SIA data~\cite{Hirai:2007cx}. The aforementioned analyses are carried out at next-to-leading order (NLO) in perturbative QCD.
Furthermore, there also exist determinations of FFs at next-to-next-to-leading order (NNLO) with SIA data only~\cite{Bertone:2017tyb,
Soleymaninia:2018uiv,Soleymaninia:2019sjo,Soleymaninia:2020bsq} and at approximate NNLO with SIA and SIDIS data~\cite{Borsa:2022vvp,AbdulKhalek:2022laj}.

In the present study, we focus on the fit based solely on the data from lepton colliders. There are numerous SIA measurements from various experiments and with different center of mass energies, including TASSO, TPC measurements below the $Z$-pole~\cite{TASSO:1988jma,TPCTwoGamma:1988yjh}, OPAL, ALEPH, DELPHI, and SLD at $Z$-pole~\cite{OPAL:1994zan,ALEPH:1994cbg,DELPHI:1998cgx,SLD:2003ogn}, and OPAL and DELPHI above the $Z$-pole~\cite{OPAL:2002isf,DELPHI:2000ahn}.
They measured the production of $\pi^{\pm}$, $K^{\pm}$ and $p/\bar p$ separately. It is widely acknowledged that these measurements from SIA are not sensitive to the separation between quark and anti-quark FFs. Furthermore, the gluon FF is poorly constrained by SIA measurements.
Consequently, a significant number of pertinent studies in the determination of light charged hadron FFs typically incorporate the data from SIDIS and pp collisions to enhance the differentiation between distinct quark and anti-quark flavors.

With the construction and operation of future lepton colliders, such as the International Linear Collider (ILC)~\cite{Behnke:2013xla,Bambade:2019fyw,ILCInternationalDevelopmentTeam:2022izu}, the Circular Electron Position Collider (CEPC)~\cite{CEPCStudyGroup:2018rmc,CEPCStudyGroup:2018ghi,CEPCStudyGroup:2023quu} and the Future Circular Collider (FCC-$ee$)~\cite{FCC:2018byv,FCC:2018evy,FCC:2018vvp}, it will be possible to have well-constrained FFs
based solely on comprehensive data sets from future lepton colliders~\cite{Liang:2023yyi,Zhu:2023xpk}. In comparison to previous lepton colliders, these colliders can operate at higher collider energies and measure a variety of observables. The CEPC and FCC-$ee$ operate primarily at energies of $91\,\GeV$, $160\,\GeV$, $240\,\GeV$ and $360\,\GeV$ to perform precision measurements on the properties of the Higgs boson, $W$ and $Z$ bosons, while the ILC runs initially at $500\,\GeV$, then at $350\,\GeV$ and $250\,\GeV$. The running scenarios at a wide range of center-of-mass energy will provide a substantial quantity of experimental data, which can be employed to constrain the FFs. For instance, the data from the decay of the Higgs boson to gluons can be used to constrain the gluon FF. Furthermore,
the large number of events at future lepton colliders will generate will lead to a reduction in both the statistical and systematic uncertainties.

In this study, we assess the constraining power of measurements from future lepton colliders on the FFs to light charged hadrons from quarks and gluon\footnote{Before our current paper, another work~\cite{Aschenauer:2019kzf} presents a quantitative assessment of the impact a future Electron-Ion Collider would have in the determination of parton-to-hadron fragmentation functions through semi-inclusive deep-inelastic electron-proton scattering data.}. We first generate pseudo--data for a number of FFs-sensitive processes at future lepton colliders.
Subsequently, based on the NPC23 set obtained with a global
analysis to current world data, we carry out an analysis of FFs at NLO to quantify the constraints of pseudo--data on the FFs.
We then conduct various alternative fits to assess the impact of individual data sets or higher-order QCD corrections on the extracted FFs. Finally, we discuss the impact of three-jet production at electron-positron colliders on the determination of FFs.
Note that, in this work, we focus mainly on the analysis carried out at NLO in QCD.

This paper is organized as follows.
In Section~\ref{sec:data} we describe the features of various processes used to generate the pseudo--data.
In Section~\ref{sec:Results and discussions}
we quantify the constraints to the FFs from these pseudo--data using the fit of FFs.
The summary comes in Section~\ref{sec:summary}.
We briefly introduce how to conduct NNLO calculations for SIA and SIDIS, as well as NLO calculations for hadron multiplicities in three jets production in the appendix, as implemented in FMNLO~\cite{Liu:2023fsq} and used in this study.

\section{Pseudo--data generation}
\label{sec:data}
In this section, we present various processes at future lepton colliders for which pseudo--data have been generated. Details on the experimental binning and kinematic cuts are provided. Additionally, the baseline measurement from SLD at $Z$--pole~\cite{SLD:2003ogn} is described, which is used to model the experimental systematic uncertainties expected in future lepton colliders.
\subsection{FFs--sensitive processes}
\label{sec:FFs--sensitive processes}
This study begins with an examination of various processes at lepton colliders that will be employed in the generation of pseudo--data.
The corresponding parton-level processes include the $q\bar{q}$ production with $q$ representing all quark flavors except for the top quark at various center-of-mass energies, $W^-W^{+,*}$ production at $\sqrt{s}\sim 2\,m_W$ together with the $W^{+,*}$ boson decaying into quark pairs, and the three hadronic decay channels of the Higgs boson. The off-shell and on-shell $W$ boson contributions are included in the $W^{+,*}$ boson. Furthermore, our analysis encompasses the heavy-flavor (either the charm or bottom quark) tagged measurements in electron-positron annihilation. The heavy-flavor tagged measurements provide crucial inputs for the determination of FFs from heavy quarks, as will be demonstrated in subsequent sections.

For the Higgs boson decay processes, we generate pseudo--data for differential distribution in the energy fraction $x_h=2E_h/\sqrt{s}$ carried by $\pi^\pm,\, K^\pm$ and $p/\bar{p}$ respectively.
Future lepton colliders, which are designed as Higgs factories, will produce a vast number of Higgs bosons.
We assume the Higgs boson are produced at a center-of-mass energy of $240$ or $250 \,\GeV$. It is known that the branching ratios of the Higgs boson decaying into bottom quark-antiquark pairs, charm quark-antiquark pairs, and gluons are $57.7\%, 2.91\%$ and $8.57\%$, respectively~\cite{LHCHiggsCrossSectionWorkingGroup:2013rie,LHCHiggsCrossSectionWorkingGroup:2016ypw}. Consequently, the expected number of events for the three hadronic decay modes can be readily derived, which can then be employed to calculate the statistical errors.

\begin{figure}[h!]
	\centering
	\includegraphics[width=0.9\textwidth]{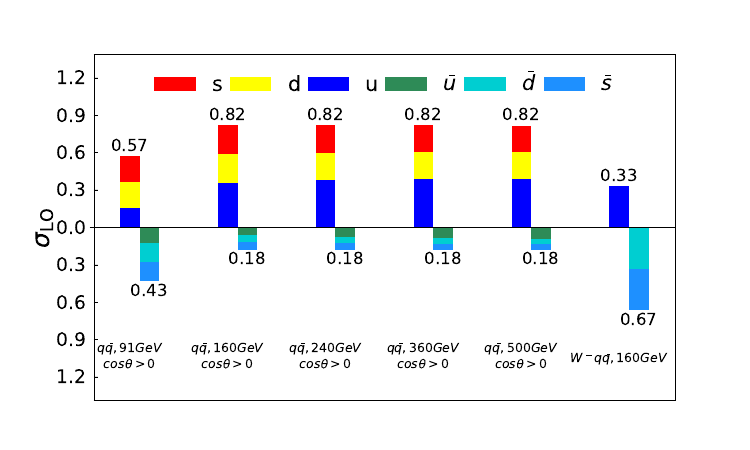}
	\vspace{-2ex}
	\caption{\label{figure:composition}
 LO cross section for various light quarks and antiquarks production at various processes. The total cross sections for each process are normalized to 1.}
\end{figure}

For the $q\bar{q}$ production at various center-of-mass energies, we focus on the distribution of the energy fraction carried by the identified hadron $\pi^+$, $\pi^-$, $K^+$, $K^-$, $p$ and $\bar{p}$ separately. The angle of emission of the identified hadron with respect to the electron beam direction in the center-of-mass frame is required to satisfy $\text{cos}(\theta)>0$.
These measurements are of great utility in the differentiation between various light quarks ($u,\,d$ and $s$ quark) and antiquark flavors.
Figure~\ref{figure:composition} illustrates the leading order (LO) cross section for the production of various light quarks and anti-quarks in different processes, with the total cross section for each process normalized to 1. For the $q\bar{q}$ production, it can be observed that the contributions from the quark and the corresponding anti-quark are significantly different at different center-of-mass energies, which is useful for the separation of light quark and anti-quark FFs. Furthermore, it can be observed that the contributions from the up-quark and the down-quark are markedly different at $\sqrt{s}=91\,\GeV$ and $160\,\GeV$, with the contribution of the various quarks remaining relatively unchanged as $\sqrt{s}$ increases.
For the $W^-W^{+*}$ production at $\sqrt{s}=160\,\GeV$ that will be discussed below, it can be seen that there are only contributions from three light (anti) quarks, which is useful to constrain the $u, \bar{d}$ and $\bar{s}$ quark FFs.
Therefore, the $q\bar{q}$ samples at different energies together with the $W$-boson pair production can ensure well-constrained FFs for light quarks.
Additionally, measurements of heavy flavors tagged with the identification of the produced hadrons ($\pi^\pm,\, K^\pm$ and $p/\bar{p}$) are considered, as these provide further inputs for the determination of FFs from heavy quarks.
In contrast to the $q\bar{q}$ production, no cuts are applied on the polar angle of the produced hadron.

In considering the $W^-W^{+*}$ production process at the reaction threshold $\sqrt{s}\sim 2\,m_W$, two measurements are distinguished: from full hadronic decays and from $c$-tagged decays. Pseudo--data are generated for the distribution of the energy fraction carried by $\pi^+$, $\pi^-$, $K^+$, $K^-$, $p$ and $\bar{p}$, respectively.
It is important to note that the measurement of particle momentum spectra can be performed in the $W^{+*}$ boson rest frame. Consequently, the energy fraction is defined as $2E_h/E_{W^{+*}}$, where $E_{W^{+*}}$ is the energy of the $W^{+*}$ boson in its rest frame.

It should be noted that the list of processes considered in this work is by no means exhaustive. It is evident that many other processes can provide valuable insights into the FFs to light charged hadrons in the forthcoming lepton collider era. For instance, the impact of three-jet production on the FFs will be presented in Section~\ref{sec:Results and discussions}. Nevertheless, the set of processes considered in this work provides a sufficient snapshot of the constraining potential from future lepton colliders.

\subsection{Theory calculations and pseudo--data generation}
\label{Theory calc}
In this subsection, we will first review the factorization theorem of the cross section and fragmentation structure functions, such as in the SIA process. The SIA differential cross section at a given center-of-mass energy of $\sqrt{s}=Q$ is written as~\cite{Nason:1993xx,Webber:1994zd}
\begin{equation}
  \label{eq:2_1}
  \frac{d^2\sigma^h}{dx_h\,d\cos(\theta)} = \frac{3}{8}(1+\cos^2(\theta))\,F_T^{h}(x_h,Q^2)
  + \frac{3}{4}\sin^2(\theta)\, F_L^{h}(x_h,Q^2) + \frac{3}{4}\cos(\theta)\,
  F_A^{h}(x_h,Q^2).
\end{equation}
where $F_T^{h}(x_h,Q^2)$, $F_L^{h}(x_h,Q^2)$ and $F_A^{h}(x_h,Q^2)$ are the transverse, longitudinal and asymmetric structure functions, respectively.
These can be expressed in a factorized form of fragmentation functions $D^h_i(x_h,\mu_F^2)$ with $i=q,\bar{q},g$ and calculable coefficient functions $C^{\rm S,NS}_{k,l} (x_h,\mu_F^2)$ as follows~\cite{Collins:1989gx}:
\begin{align}
  \label{eq:2_2}
  F_k^{h}(x_h,Q^2)= &
    \sigma_{\rm tot}^{(0)}(Q^2)\,
    [D_{\rm S}^h\left(x_h,\mu_F^2\right)\otimes
    \mathbb{C}_{k,q}^{\rm S}(x_h,\mu_F^2)  + D_g^h\left(x_h,\mu_F^2\right)\otimes\mathbb{C}_{k,g}^{\rm S}(x_h,\mu_F^2)] \nonumber\\
    &
    +\sum_{p=1}^{n_f}\,\sigma_p^{(0)}(Q^2)\,D_{ {\rm NS},p}^h\left(x_h,\mu_F^2\right)\otimes
  \mathbb{C}_{k,q}^{\rm NS}(x_h,\mu_F^2), \quad {\rm for \, k=T,L}\nonumber\\
  F_A^{h}(x_h,Q^2) =& \sum_{p=1}^{n_f}
  A_p^{(0)}(Q^2)
  D_{A,p}^h(x_h,\mu_F^2) \otimes \mathbb{C}_{A,q}^{\rm NS}(x_h,\mu_F^2).
\end{align}
The NNLO coefficient functions $C^{\rm S, NS}_{k, l}$ with $k=T, L, A$ and $l=q, g$ have been calculated in Refs.~\cite{Rijken:1996vr,Rijken:1996npa,Rijken:1996ns,Mitov:2006wy,Blumlein:2006rr}.
The FFs $D^h_i(x_h,\mu_F^2)$ are related to a nonperturbative aspect of QCD, they cannot be
calculated from the first principle of QCD at present.
However, their scale dependence follows the Dokshitzer-Gribov-Lipatov-Altarelli-Parisi (DGLAP) evolution equation with time-like
splitting kernels, which can be perturbatively calculated. The time-like splitting functions have been calculated up to $\mathcal{O}(\alpha_S^3)$ in the strong coupling constant and can be found in Refs.~\cite{Mitov:2006ic,Moch:2007tx,Almasy:2011eq,Chen:2020uvt,Luo:2020epw,Ebert:2020qef}.
The renormalization scale $\mu_R$ has been set equal to the mass factorization scale $\mu_F$ for simplicity.
In Eq.~\eqref{eq:2_2}, $\sigma_p^{(0)}$, where the indices $p = 1, 2, 3, 4, 5$ stand for $u, d, s, c, b$ quark, is the LO cross section for the production of the quark $p$. $\sigma_{\rm tot}^{(0)}$ is the corresponding sum over all $n_f=5$ active flavors, $\sigma ^{(0)}_{\rm tot} = \sum_{p=1}^{n_f} \sigma ^{(0)}_p$, while $A_p^{(0)}$ is the asymmetry factor to which only non-singlet channels contribute. In Eq.~\eqref{eq:2_2}, the symbol $\otimes$ denotes the standard convolution integral, defined as

\begin{equation}\label{convolution}
f(z) \otimes g(z)
=\int^1_0 dx \int^1_0 dy f(x) g(y) \delta(z - xy) \,,
\end{equation}
and we have defined the singlet (S) and non-singlet (NS, A) combinations of quark fragmentation functions. These are given by
\begin{align}
\label{singlet}
&D_S^h(x_h, \mu_F^2) = \frac{1}{n_f}
\sum_{p=1}^{n_f} [D^h_p(x_h, \mu_F^2) + D^h_{\bar{p}}(x_h, \mu_F^2)] \,,\\
&D_{{\rm NS}, p}^h(x_h, \mu_F^2) =
D^h_p(x_h, \mu_F^2) + D^h_{\bar{p}}(x_h, \mu_F^2) - D^h_{\rm S}(x_h, \mu_F^2)\,,\\
&D_{\rm A, p}^h(x_h, \mu_F^2) =
D^h_p(x_h, \mu_F^2) - D^h_{\bar{p}}(x_h, \mu_F^2).
\end{align}
As in the case of SIA, the differential decay width for the $H\to g g$ process in the heavy-top limit quark is given by
\begin{equation}
  \label{eq:2_3}
  \frac{d\Gamma^h}{dx_h} = \Gamma_0\,\left(D_{\rm S}^h\left(x_h,\mu_F^2\right)\otimes
    \mathbb{C}_{q}(x_h,\mu_F^2)  + D_g^h\left(x_h,\mu_F^2\right)\otimes\mathbb{C}_{g}(x_h,\mu_F^2)\right).
\end{equation}
where $\Gamma_0$ is the LO decay width in the heavy-top quark limit and the coefficient functions $\mathbb{C}_{q/g}(x_h,\mu_F^2)$ have been calculated up to NNLO~\cite{Soar:2009yh,Almasy:2011eq}.
For the production of $W$ boson pairs and the decay of the Higgs boson to heavy quarks (c and b),
their NNLO coefficient functions are not yet known. Furthermore, for the present study based on the pseudo--data, we are mostly interested in the reduction of experimental uncertainties of FFs.
Thus, in this work, we focus on the analysis carried out at NLO in QCD.

The NLO predictions of the described above processes are calculated with the FMNLO program~\cite{Liu:2023fsq}, which enables fast convolution with FFs using stored grids of hard coefficient functions. The central values of the pseudo--data initially coincide with the corresponding prediction obtained from NLO calculation with the $\rm NPC23$ FFs~\cite{Gao:2024nkz,Gao:2024dbv} as input.
In Appendix~\ref{Appendix:HKNS}, we also use the
HKNS FFs~\cite{Hirai:2007cx}, which are extracted from SIA data only, to generate pseudo--data and perform the extraction of the FFs.
Subsequently, the central values are subjected to fluctuations in accordance with the corresponding experimental uncertainties.
This implies that, by construction, one should find $\chi^2/N_{\rm pt}\simeq 1$ from the fit to the pseudo--data.
Statistical uncertainties are evaluated from the expected number of events per bin, while systematic uncertainties at future lepton colliders are not yet known.
The performance of future lepton colliders and detectors, in principle, should be improved compared to the last generation, like SLD. For simplicity and to be conservative, we simply assume the systematic uncertainties are the same as those in the measurements from SLD at $Z$--pole~\cite{SLD:2003ogn}.
The identified $\pi^{\pm}$, $K^{\pm}$ and $p/\bar{p}$ production from hadronic decays of $Z$ boson are measured by SLD. In addition to flavor-inclusive $Z$ boson decays, measurements are available for $Z$ boson decays into light ($u, d, s$), $c$, and $b$ primary flavors. Table~\ref{tab:binning} presents the binning and the corresponding uncorrelated systematic (relative) uncertainties for the charged kaon distribution of the scaled momentum, as measured by SLD. The correlated systematic (relative) uncertainty is $1\%$.
Table~\ref{tab:binning} reveals that the uncorrelated systematic uncertainty ranges from $1.8\%$ to $32\%$.
We choose not to generate the central values of pseudo--data
from general-purpose event generators. The main reason is that they often rely on phenomenological models for hadronization, such as the Lund string model or the cluster model. In contrast, our theoretical framework is based on QCD collinear factorization. Especially, we carry out the NLO analysis which uses two-loop time-like splitting kernels for the evolution of FFs. On the other hand, for almost all event generators on the markets, the parton shower is carried out only at LO, which corresponds to one-loop DGLAP evolution.

\begin{table}
\begin{center}
 \begin{tabular}{|c|c|c|c|c|c|} \hline
 $x_h$ Range  &  $\delta^{\rm exp}_{{\rm sys}}$ [$\%$] & $x_h$ Range & $\delta^{\rm exp}_{{\rm sys}}$ [$\%$]  & $x_h$ Range & $\delta^{\rm exp}_{{\rm sys}}$ [$\%$]  \\ \hline
0.014--0.016 & 32  & 0.077--0.082  & 6.0 & 0.274--0.296  & 1.9 \\
0.016--0.022 & 7.3  & 0.082--0.088  & 6.0  & 0.296--0.318  & 1.8 \\
0.022--0.027 & 3.7 & 0.088--0.099  & 5.9  & 0.318--0.351  & 1.9 \\
0.027--0.033 & 2.7  & 0.099--0.110  & 5.7  & 0.351--0.384  & 2.3 \\
0.033--0.038 & 2.4  & 0.110--0.121  & 5.5  &0.384--0.417   & 2.6 \\
0.038--0.044 & 2.5  & 0.121--0.143  & 6.4  &0.417--0.450   & 3.2 \\
0.044--0.049 & 3.0  & 0.143--0.164  & 9.6  &0.450--0.482   & 3.4 \\
0.049--0.055 & 3.0 & 0.164--0.186  & 15  &0.482--0.526   & 3.8 \\
0.055--0.060 &3.7& 0.186--0.208  & 15  &0.526--0.570   & 4.3 \\
0.060--0.066 & 4.4 & 0.208--0.230  & 6.6  &0.570--0.658   & 4.9 \\
0.066--0.071 & 5.3 & 0.230--0.252  & 3.3  &0.658--0.768   &5.6 \\
0.071--0.077 & 5.5 & 0.252--0.274  & 2.3  &0.768--1.000   &14
 \\[.1cm] \hline
 \end{tabular}
\caption{\label{tab:binning} The binning and the corresponding uncorrelated systematic (relative) uncertainties for the charged kaon distribution of the scaled momentum measured by SLD.}
\end{center}
\end{table}

We now proceed to present further details on pseudo--data generation.
Assuming $\sigma_i^{\rm th}$ is the theoretical cross section for bin $i$ of a given process,
the central value of the pseudo--data $\sigma_i^{\rm exp}$ is constructed as
\begin{align}
\label{eq:pseudodataGen}
\sigma_i^{\rm exp}  = \sigma_i^{\rm th}\times (1+r_i\cdot\delta_{{\rm tot},\,i}^{\rm exp}+s\cdot\delta_{\rm cor}^{\rm exp})
\end{align}
where $r_i$ and $s$ are univariate Gaussian random numbers, while $\delta^{\rm exp}_{{\rm tot}, i}$ denotes the uncorrelated experimental uncertainty associated with this specific bin. Furthermore, $\delta^{\rm exp}_{\rm cor}$ represents correlated systematic uncertainty, which is
assumed to be $1\%$ across bins of each of the pseudo--data.

In Eq.~(\ref{eq:pseudodataGen}), the uncorrelated experimental
uncertainty $\delta^{\rm exp}_{{\rm tot},i}$
includes both statistical and systematic uncertainties and is defined as
\begin{align}
\label{eq:totalExpError}
\delta^{\rm exp}_{{\rm tot},i} \equiv \left((\delta^{\rm exp}_{{\rm stat},i})^2+(\delta^{\rm exp}_{{\rm sys},i})^2\right)^{1/2}.
\end{align}
In this expression, the relative statistical uncertainty $\delta^{\rm exp}_{{\rm stat},i}$ is
computed as
\begin{align}
\delta^{\rm exp}_{{\rm stat},i}=(N_{{\rm ev},i})^{-1/2},
\end{align}
where $N_{{\rm ev},i}=\sigma_i^{\rm th} \times \mathcal{L}$ is the expected number of events in bin $i$ with $\mathcal{L}$ being integrated luminosity.
In Eq.~(\ref{eq:totalExpError}), $\delta^{\rm exp}_{{\rm sys}, i}$ indicates the uncorrelated systematic uncertainty of bin $i$, which is derived from the reference SLD measurement at $Z$--pole as summarized in Table~\ref{tab:binning}. Note we use the same bininig of the scaled momentum as the SLD measurements as well.

\begin{table}[t!]
\centering
\scalebox{0.85}{
\begin{tabular}{|c|c|c|c|c|c|c|c|}
\hline
\multicolumn{8}{|c|}{$e^+e^- \,\text{annihilation}$} \\\hline
\multirow{2}{*}{$\sqrt{s}\,(\GeV)$} & \multicolumn{3}{|c|}{$\rm luminosity \,(ab^{-1})$} & \multirow{2}{*}{$\text{final state}$} & \multirow{2}{*}{$\text{kinematic cuts}$} & \multirow{2}{*}{$\text{hadrons}$} & \multirow{2}{*}{$N_{\rm pt}$} \\ \cline{2-4}
&CEPC & FCC-$ee$& ILC &  & & &  \\ \cline{1-8}
\multirow{2}{*}{\makecell{$91.2$}}
 & \multirow{2}{*}{$60$}& \multirow{2}{*}{$150$}& \multirow{2}{*}{$\text{-}$}& $q\bar{q}$ & $\text{cos}(\theta)>0$ &  $h^{+,-}$ & $132$ \\  \cline{5-8}
 & & & &$c\bar{c}/b\bar{b}$ & $\text{-}$ & $h^{\pm}$ & $65$\\ \cline{1-8}

\multirow{2}{*}{\makecell{$160$}}
 & \multirow{2}{*}{$4.2$}& \multirow{2}{*}{$\text{-}$}& \multirow{2}{*}{$\text{-}$}& $q\bar{q}$ & $\text{cos}(\theta)>0$ &  $h^{+,-}$ & $168$ \\  \cline{5-8}
 & & & &$c\bar{c}/b\bar{b}$ & $\text{-}$ & $h^{\pm}$ & $83$\\ \cline{1-8}

\multirow{2}{*}{\makecell{$161$}}
 & \multirow{2}{*}{$\text{-}$}& \multirow{2}{*}{$10$}& \multirow{2}{*}{$\text{-}$}& $q\bar{q}$ & $\text{cos}(\theta)>0$ &  $h^{+,-}$ & $168$ \\  \cline{5-8}
 & & & &$c\bar{c}/b\bar{b}$ & $\text{-}$ & $h^{\pm}$ & $83$\\ \cline{1-8}

\multirow{2}{*}{\makecell{$240$}}
 & \multirow{2}{*}{$13$}& \multirow{2}{*}{$5$}& \multirow{2}{*}{$\text{-}$}& $q\bar{q}$ & $\text{cos}(\theta)>0$ &  $h^{+,-}$ & $186$ \\  \cline{5-8}
 & & & &$c\bar{c}/b\bar{b}$ & $\text{-}$ & $h^{\pm}$ & $92$\\ \cline{1-8}

\multirow{2}{*}{\makecell{$250$}}
 & \multirow{2}{*}{$\text{-}$}& \multirow{2}{*}{$\text{-}$}& \multirow{2}{*}{$2$}& $q\bar{q}$ & $\text{cos}(\theta)>0$ &  $h^{+,-}$ & $186$ \\  \cline{5-8}
 & & & &$c\bar{c}/b\bar{b}$ & $\text{-}$ & $h^{\pm}$ & $92$\\ \cline{1-8}

\multirow{2}{*}{\makecell{$350$}}
 & \multirow{2}{*}{$\text{-}$}& \multirow{2}{*}{$0.2$}& \multirow{2}{*}{$0.2$}& $q\bar{q}$ & $\text{cos}(\theta)>0$ &  $h^{+,-}$ & $198$ \\  \cline{5-8}
 & & & &$c\bar{c}/b\bar{b}$ & $\text{-}$ &$h^{\pm}$ & $98$\\ \cline{1-8}

\multirow{2}{*}{\makecell{$360$}}
 & \multirow{2}{*}{$0.65$}& \multirow{2}{*}{$\text{-}$}& \multirow{2}{*}{$\text{-}$}& $q\bar{q}$ & $\text{cos}(\theta)>0$ &  $h^{+,-}$ & $198$ \\  \cline{5-8}
 & & & &$c\bar{c}/b\bar{b}$ & $\text{-}$ & $h^{\pm}$ & $98$\\ \cline{1-8}

\multirow{2}{*}{\makecell{$365$}}
 & \multirow{2}{*}{$\text{-}$}& \multirow{2}{*}{$1.5$}& \multirow{2}{*}{$\text{-}$}& $q\bar{q}$ & $\text{cos}(\theta)>0$ &  $h^{+,-}$ & $198$ \\  \cline{5-8}
 & & & &$c\bar{c}/b\bar{b}$ &$\text{-}$ & $h^{\pm}$ & $98$\\ \cline{1-8}

\multirow{2}{*}{\makecell{$500$}}
 & \multirow{2}{*}{$\text{-}$}& \multirow{2}{*}{$\text{-}$}& \multirow{2}{*}{$4$}& $q\bar{q}$ & $\text{cos}(\theta)>0$ &  $h^{+,-}$ & $198$ \\  \cline{5-8}
 & & & &$c\bar{c}/b\bar{b}$ & $\text{-}$ & $h^{\pm}$ & $98$\\ \cline{1-8}

\multicolumn{8}{|c|}{$\text{W boson decay channels}$} \\\hline
\multirow{2}{*}{$\sqrt{s}\,(\GeV)$} & \multicolumn{3}{|c|}{$\text{\# events (million)}$} & \multirow{2}{*}{$\text{final state}$} & \multirow{2}{*}{$\text{kinematic cuts}$} & \multirow{2}{*}{$\text{hadrons}$} & \multirow{2}{*}{$N_{\rm pt}$} \\ \cline{2-4}
&CEPC & FCC-$ee$& ILC &  & & &  \\ \cline{1-8}
\multirow{2}{*}{\makecell{$80.419$}}
 & $116$& $68$& $62$& $W^-W^{+*}\to W^-q\bar{q}$ & \multirow{2}{*}{$\text{-}$} &  \multirow{2}{*}{$h^{+,-}$} & \multirow{2}{*}{$120$}  \\ \cline{2-5}
 &$58$& $34$& $31$ & $W^-W^{+*}\to W^-c\bar{s}$ &  &  & \\ \cline{1-8}

\multicolumn{8}{|c|}{$\text{Higgs boson decay channels}$} \\\hline
\multirow{2}{*}{$\sqrt{s}\,(\GeV)$} & \multicolumn{3}{|c|}{$\text{\# events (million)}$} & \multirow{2}{*}{$\text{final state}$} & \multirow{2}{*}{$\text{kinematic cuts}$} & \multirow{2}{*}{$\text{hadrons}$} & \multirow{2}{*}{$N_{\rm pt}$} \\ \cline{2-4}
&CEPC & FCC-$ee$& ILC &  & & &  \\ \cline{1-8}
\multirow{3}{*}{\makecell{$125$}}
 & $0.23$& $0.09$& $0.07$& $gg$ & \multirow{3}{*}{$\text{-}$} &  \multirow{3}{*}{$h^{\pm}$} & \multirow{3}{*}{$77$}  \\ \cline{2-5}
 &$0.08$& $0.03$& $0.02$ & $c\bar{c}$ &  &  & \\ \cline{2-5}
 &$1.53$& $0.59$& $0.47$ & $b\bar{b}$ &  & & \\ \cline{1-8}

\end{tabular}
}
\caption{
A summary of the main features of CEPC, FCC-$ee$, and ILC pseudo--data generated for the present study.
For each process, we indicate the
center-of-mass energy, luminosity (number of events), final states, kinematic cuts, the identified hadrons, and the number of data points after selections. Here $h^{\pm}$ and $h^{+,-}$ denote $(\pi^\pm,K^\pm,p/\bar{p})$ and $(\pi^+,\pi^-,K^+,K^-,p,\bar{p})$, respectively.}
\label{tab:PseudoData}
\end{table}

In Table~\ref{tab:PseudoData} we present a summary of the main features of the CEPC, FCC-$ee$, and ILC pseudo--data generated for the present study, respectively.
For each process, we indicate the
center-of-mass energy, luminosity (number of events), final states, kinematic cuts, the identified hadrons, and finally the number of data points after selections.
The expected integrated luminosities refer to the total luminosities over the entire run period, summed over two interaction points. Moreover, the anticipated number of events per bin for the $q\bar{q}$ production with $\cos(\theta)>0$ cut must be multiplied by two, due to the contributions from $\cos(\theta)<0$. A similar situation arises in the case of $W^-W^{+,*}$ production because of the contributions from $W^+W^{-,*}$ production.
In Table~\ref{tab:PseudoData}, we have combined the expected number of events for $W^-W^{+,*}$ production at center-of-mass energy of $160\,\GeV$ and above.

It is ﬁnally necessary to mention that we have implemented stringent selection criteria on hadron kinematics, requiring both
the hadron energy fraction $x_h > 0.01$ and the hadron energy $E_h > 4\,\GeV$, to ensure the validity of leading-twist factorization and the convergence of perturbative calculations. The hadron energy is measured in the center-of-mass frame and the $W^{+,*}$ or Higgs boson rest frame for the quark pair production and $W^-W^{+,*}$ production or Higgs boson decay, respectively.

\subsection{Theoretical setup and fit of FFs}
The parameterization form of fragmentation functions to charged hadrons used at the initial scale $Q_0$ is
\begin{equation}\label{eq:para}
xD_{h/i}\left(x, Q_{0}\right)=
x^{\alpha_i^h}(1-x)^{\beta_i^h} {\rm exp}\left(\sum_{n=0}^m a_{i,n}^h(\sqrt{x})^n\right),
\end{equation}
where $\{\alpha, \beta, a_{n}\}$ are free parameters in the fit, and $i$ and $h$ indicate the flavor of the parton and the hadron, respectively.
We choose $Q_0=5\,\GeV$  and use a zero-mass scheme for heavy quarks.
One advantage of the above parametrization form is that the fragmentation functions are positively
defined, thus no additional positivity constraints need to be applied. The total number of free parameters is 63 for $\pi^+, K^+$ and $p$ combined. Further details can be found in Ref.~\cite{Gao:2024dbv}. Furthermore, the FFs of negative-charged
hadrons are related to those of positive-charged
hadrons via charge conjugation.
The fragmentation functions are evolved to higher scales using two-loop
time-like splitting kernels to be consistent with the NLO analysis.
The splitting functions were calculated in Refs.~\cite{Curci:1980uw,Furmanski:1980cm,Floratos:1981hs,Kalinowski:1980ju,Stratmann:1996hn} and are
implemented in HOPPET~\cite{Salam:2008qg,Salam:2008sz} which we use in the analysis.

The quality of the agreement between new experimental measurements and the
corresponding theoretical predictions for a given set of
fragmentation parameters is quantified by the log-likelihood function ($\chi^2$), which is given by~\cite{Gao:2017yyd}
\begin{equation}
  \chi^2 (\{\alpha,\beta,a_n\}, \{\lambda\}) = \sum_{k = 1}^{N_{\rm pt}} \frac{1}{s_k^2}
  \left( D_k - T_k - \sum_{\mu = 1}^{N_{\lambda}} \sigma_{k, \mu}
  \lambda_{\mu} \right)^2 + \sum_{\mu = 1}^{N_{\lambda}}
  \lambda_{\mu}^2.
\label{eq:chi2}
\end{equation}
$N_{\rm pt}$ is the number of data points, the nuisance parameters $\lambda_{\mu}$ describe sources of correlated errors, which are assumed
to follow standard normal distributions, $s^2_k$ are the total
uncorrelated uncertainties by
adding statistical and uncorrelated systematic uncertainties in quadrature,
$D_k$ is the central value of the experimental measurements, and $T_k$ is the
corresponding theoretical prediction which depends on $\{\alpha,\beta,a_n\}$.
%
The correlated uncertainty $\sigma_{k, \mu}$ quantifies the sensitivity of the $k$-th measurement to the $\mu$-th correlated error source.
%

%

%
By minimizing $\chi^2 (\{\alpha,\beta,a_n\}, \{ \lambda \})$ with respect to the
nuisance parameters, we get the profiled $\chi^2$ function
\begin{equation}
  \chi^2 (\{\alpha,\beta,a_n\},\{\hat{\lambda}\}) = \sum_{i, j = 1}^{N_{\rm pt}} (T_i - D_i) [{\rm cov}^{-
  1}]_{ij}  (T_j - D_j),
\end{equation}
where ${\rm cov}^{- 1}$ is the inverse of the covariance matrix
\begin{equation}
\label{eq:covmat}
  (\mathrm{cov})_{ij} \equiv s_i^2 \delta_{ij} + \sum_{\mu =
  1}^{N_{\lambda}} \sigma_{i, \mu} \sigma_{j, \mu},\quad (\mathrm{cov}^{-1})_{ij} \equiv \frac{\delta_{ij}}{s_i^2} - \sum_{\mu,\nu =
  1}^{N_{\lambda}} \frac{\sigma_{i, \mu}}{s_i^2} A^{-1}_{\mu\nu}\frac{\sigma_{j, \nu}}{s_j^2}.
\end{equation}
The matrix $A_{\mu\nu}$ is defined as
\begin{equation}
A_{\mu\nu} = \delta_{\mu\nu}+\sum_{k=1}^{N_{\rm pt}}\frac{\sigma_{k, \mu}\sigma_{k, \nu}}{s_k^2}.
\end{equation}
We include theoretical uncertainties into the covariance matrix of Eq.~(\ref{eq:covmat}) by default, assuming these to be fully correlated among different data points in each pseudo--data.
%
The theoretical uncertainty $\sigma_{j,\mu}$ is estimated by the half-width of the envelope
of theoretical predictions based on 9 scale combinations, namely $\mu_F/\mu_{F,0}= \{1/2, 1, 2\}$ and $\mu_R/\mu_{R,0}=\{1/2, 1, 2\}$. The central values of these scales, namely $\mu_{F,0}$ and $\mu_{R,0}$, are set to the center-of-mass energy, $\sqrt{s}$, for both $q\bar{q}$ production and the Higgs boson decay processes. In the case of $W^-W^{+,*}$ production, the value of $\mu_{R,0}$ is set to the mass of the $W$ boson, $m_W$, while the value of $\mu_{F,0}$ is set to the invariant mass of the $W^{+,*}$ boson.

The best-fit fragmentation parameters are determined by minimizing the $\chi^2$ and then further validated through a
series of profile scans on each of those parameters. These parameter space scans are conducted using the $\text{MINUIT}$~\cite{James:1975dr} program. We apply a criterion of $\Delta\chi^2\sim1$ to determine parameter uncertainties.
Additionally, we employ the iterative Hessian approach~\cite{Pumplin:2000vx} to generate error sets of fragmentation functions, which
can then be used to propagate parameter uncertainties to physical observables.

\begin{figure}[h!]
	\centering
        \vspace{-.3cm}
	\includegraphics[width=\textwidth]{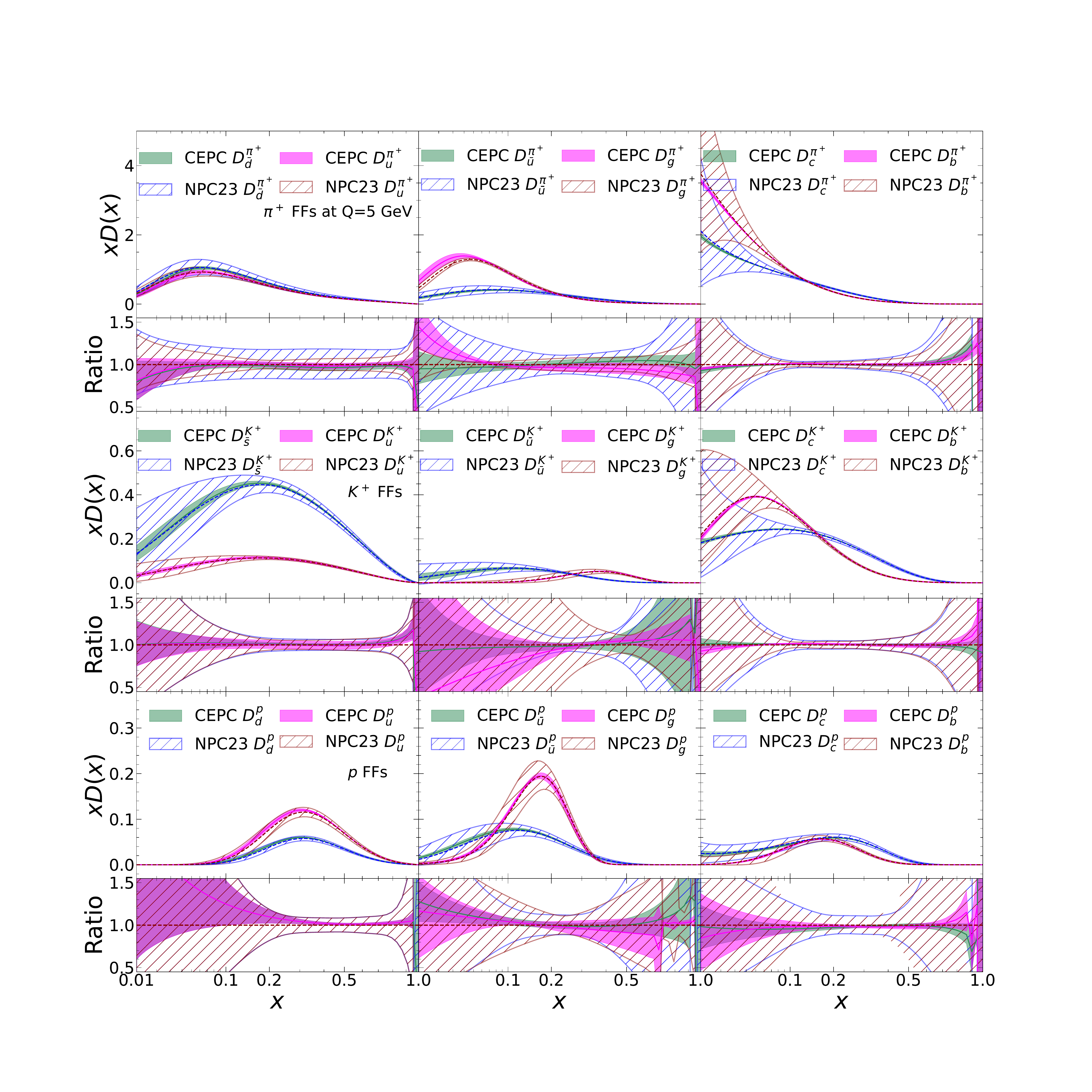}
	\vspace{-1.5cm}
	\caption{\label{figure:comNPC23_asym_p_5GeV}
A comparison of our fragmentation functions to those of NPC23 on the diverse outcomes of parton fragmentation, including $\pi^+$, $K^+$, and $p$, at an energy of $5 \,\GeV$. The colored band represents the uncertainty estimated with the Hessian Method at the 68\% confidence level, with their ratios normalized to the central value of the baseline from the NPC23 study displayed in the lower panel of each subplot.}
\end{figure}

\section{Results and discussions}
\label{sec:Results and discussions}
In this section, we utilize the sets of pseudo--data enumerated in Table~\ref{tab:PseudoData} to fit various parton FFs to three light charged hadrons and assess the extent to which the pseudo--data constrain the FFs. Subsequently, we conduct an alternative fit to evaluate the impact of theoretical uncertainties on the extracted FFs. Finally, we discuss the potential constraints from measurements of three-jet production.
\subsection{The constraints to FFs from future lepton colliders}
We ﬁrst present the results for the CEPC. The total $\chi^2$ is 1659.2 for a total number of data points of 1831, resulting in $\chi^2/N_{\rm pt} = 0.91$, which is explained in subsection~\ref{Theory calc}. A comparison of the baseline NPC23 FFs with those extracted with CEPC pseudo--data at the scale $Q_0 = 5 \,\GeV$ is presented in Figure~\ref{figure:comNPC23_asym_p_5GeV}. For simplicity, only the FFs for various partons fragmenting into $\pi^+, K^+$, and $p$ are shown. The FFs for negative charge hadrons are obtained through charge conjugation. In each row, the sub-figures in the left column show the FFs of the constituent quarks to the charged hadrons, namely the $u$ and $\bar{d}$ quarks to $\pi^+$, the $u$ and $\bar{s}$ quarks to $K^+$, and the $u$ and $d$ quarks to $p$. The remaining sub-figures correspond to the FFs of remaining light quarks and gluon, as well as $c$ and $b$ quarks to the identified hadrons, respectively. Both the absolute values of momentum fraction times the FFs and their ratios, normalized to the central values of the $\rm NPC23$ baseline, are shown. In this comparison, the colored band represents the associated uncertainty estimated with the Hessian method at the 68\% confidence level (C.L.).

As anticipated, the central values of $\rm NPC23$ and our fit exhibit a good agreement, spanning a broad range of $x$ values in all cases.
It can also be observed that at small and large $x$ there are significant deviations in the central values.
The differences at small $x$ are due to the insufficient number of data points available at small $x$ after applying the kinematic selection, while the deviations at large $x$ can be attributed to the fact that there are not enough statistics in this region. However, the differences are within the Hessian errors of the FFs.
These can be clearly seen in the lower panel of each figure.
For the quarks fragmenting into light charged hadrons, we observe a marked reduction in the FFs uncertainties. This is of particular significance for the $c$ and $b$ quarks, for the reason that a considerable number of measurements from heavy-flavor tagged hadronic events in $e^+e^-$ collisions have strong constraints to heavy quark FFs.
In the cases of the constituent quarks and sea quarks, there is a reduction in the FFs uncertainties across a wide range of $x$.
The primary reason for this improvement is the increased statistics and the diverse measurements across a wide range of collision energies, from the $Z$-pole to $360\,\GeV$, which will be available at the CEPC.

In the case of the gluon, it can be observed that CEPC can significantly reduce the uncertainties of FFs for the gluon fragmenting into $K^+$ and $p$ across the entire kinematic region, compared to the NPC23 results.
The FFs to $\pi^+$ from gluon have reduced uncertainties at $x>0.08$.
These results highlight the importance of the measurements for the $H\to gg$ channel which represent a unique handle on the poorly known gluon FF to three light charged hadrons in our work.
Furthermore, we observe a slight increase of uncertainties for the gluon FF to $\pi^+$ in the small $x$ region. This can be attributed to the insufficient number of data points available for the determination of gluon FFs to $\pi^+$ after data selection in the small $x$ region.

\begin{figure}[h!]
	\centering
 \vspace{-.3cm}
	\includegraphics[width=\textwidth]{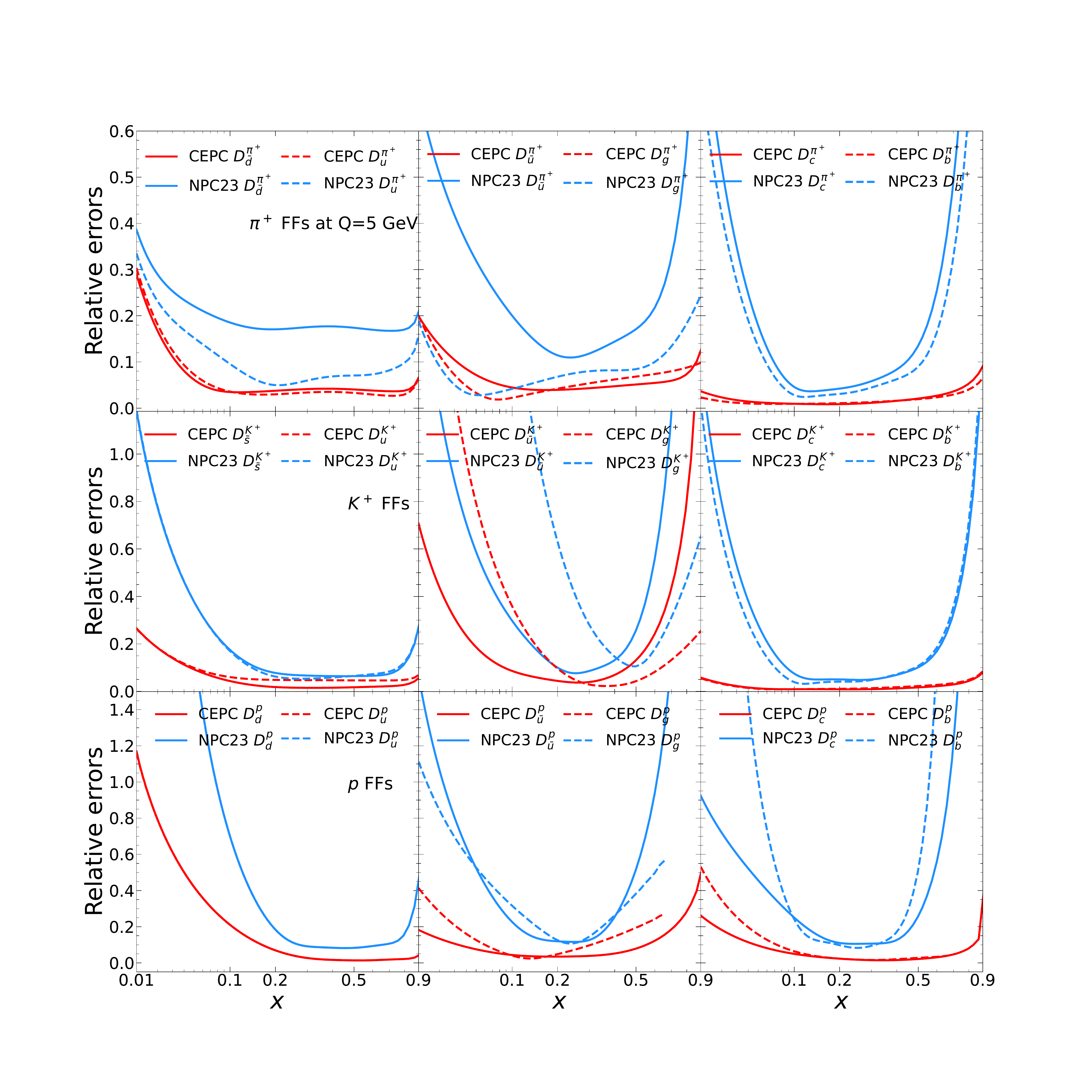}
	\vspace{-1.5cm}
    \caption{
\label{figure:comNPC23_sym_p_5GeV}A comparison of the symmetric relative errors between the $\pi^+$, $K^+$ and $p$ fragmentation functions from the CEPC and the $\rm NPC23$ sets.}
\end{figure}

In Figure~\ref{figure:comNPC23_sym_p_5GeV}, we present a comparison of the relative errors between the $\pi^+$, $K^+$ and $p$ FFs from CEPC and $\rm NPC23$ at the scale $Q_0 = 5 \,\GeV$. In the case of $\pi^+$ FFs, the relative errors observed in the CEPC study are considerably smaller than those observed in the $\rm NPC23$ FFs, with the exception of the FFs to $\pi^+$ from gluon in the small $x$ region, which has been previously explained. For instance, the relative errors for FFs of the constituent quark $\bar{d}$ to $\pi^+$ are reduced by approximately a factor of five, from approximately 20\% to a few percents, in a wide region of $x$. Furthermore, we observe a pronounced reduction in the relative errors of the FFs to $K^+$ and $p$. This is particularly significant in the small and large $x$ regions. The relative errors for the constituent quarks $u$ and $d$ fragmenting into $p$ are identical for the reason that we have assumed flavor symmetry between the constituent quarks $D_u^p(x, Q_0)=2\, D_d^p(x, Q_0)$ at the starting scale $Q_0$.

\begin{figure}[h!]
	\centering
 \vspace{-.3cm}
	\includegraphics[width=\textwidth]{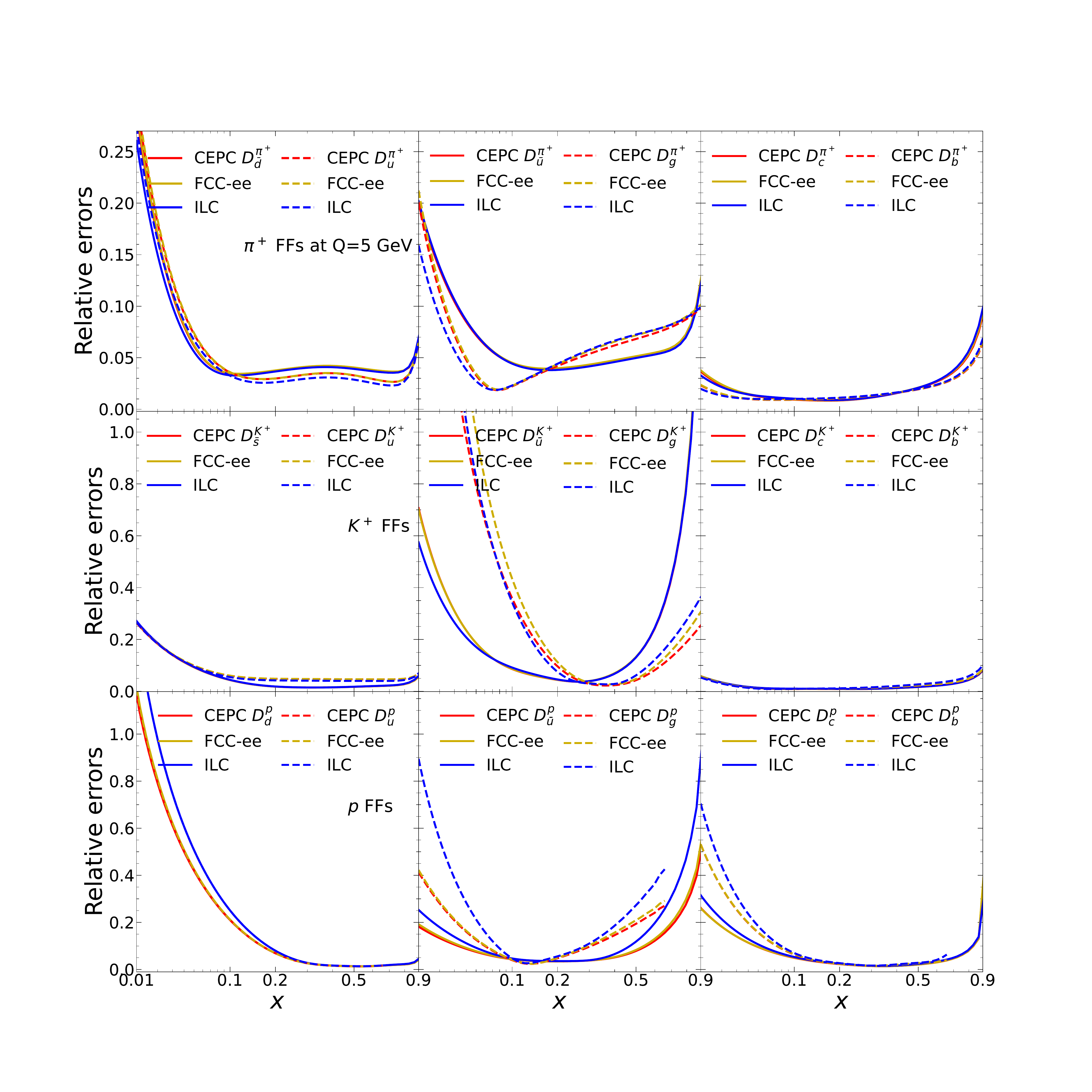}
	\vspace{-1.5cm}
    \caption{
\label{figure:lepton_p_5GeV}A comparison of the relative errors among the FFs obtained from CEPC, FCC-$ee$ and ILC measurements, respectively, at the scale $Q_0 = 5 \,\GeV$.}
\end{figure}

We now present in Figure~\ref{figure:lepton_p_5GeV} a comparison of the relative errors among the FFs obtained from CEPC, FCC-$ee$, and ILC pseudo--data, at the scale $Q_0 = 5 \,\GeV$.
It can be observed that the pseudo--data from the CEPC and FCC-$ee$ have a comparable impact on uncertainties of FFs.
This is because the two colliders have similar parameters of luminosities and center-of-mass energies. In the case of the ILC, the constraints on the $\pi^+$ and $K^+$ FFs are in general comparable to those observed in the CEPC case. However, there are stronger constraints on the $\pi^+$ FFs from gluon and $K^+$ FFs from the $\bar{u}$ quark in the small $x$ region, in comparison to the CEPC.
This is because the ILC has a higher center-of-mass energy and therefore there are more data points from the ILC at small $x$ after applying the selection criteria on the hadron energy $E_h > 4 \rm GeV$. For the $p$ FFs, it can be observed that the relative errors of the FFs extracted from ILC measurements are larger than those from
 the CEPC in the entire region of $x$.
This is because the ILC has larger statistical errors than those of the CEPC or the FCC-$ee$ in the pseudo--data of proton production.

\subsection{Alternative fit}
\label{sec:Alternative fit}
In this subsection, we utilize the FFs from the CEPC as a case study to investigate the impact of individual data sets and theoretical uncertainties on the extracted FFs. By comparing these alternative fits with the baseline fit derived from the complete data sets presented in Table~\ref{tab:PseudoData}, we aim to identify potential constraints from different data sets and the influence of higher-order QCD corrections.
To evaluate the impact of particular data sets on distinct fragmentation processes, alternative fits are conducted by systematically excluding a specific group of data sets at a time and re-fitting the FFs at NLO. Throughout the fits, all other variables are kept unchanged compared to the baseline fit, including the kinematic cuts and the treatment of theoretical uncertainties.

\begin{figure}[t!]
	\centering
 \vspace{-.3cm}
	\includegraphics[width=\textwidth]{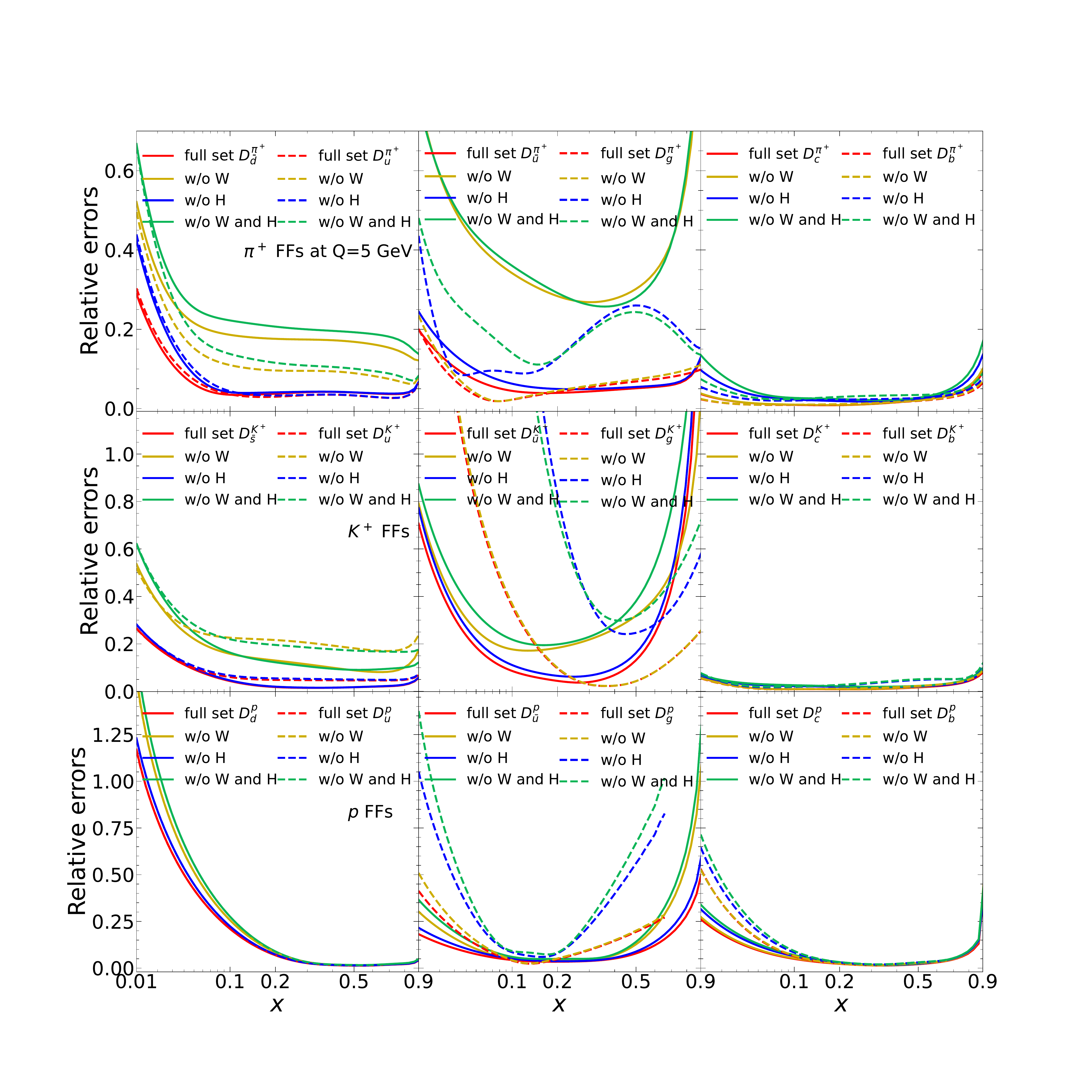}
	\vspace{-1.5cm}
	\caption{\label{figure:comwo_sym_p_5GeV}
 A comparison of the relative errors on the $\pi^+$, $K^+$ and $p$ FFs from the baseline fit and a refitted version by excluding a specific group of data sets presented in Table~\ref{tab:PseudoData}.}
\end{figure}
A comparison of the relative errors on the $\pi^+$, $K^+$ and $p$ FFs from the baseline fit and a refitted version, which excludes a specific group of data sets, is presented in Figure~\ref{figure:comwo_sym_p_5GeV}.
Upon the removal of the $W^-W^{+,*}$ production data sets, it becomes evident that the relative errors for the FFs of light (anti) quarks are larger than those of the baseline fit. This is because the pseudo--data from the $W^-W^{+,*}$ production is useful to constrain the $u$, $\bar{d}$ and $\bar{s}$ FFs, which has been explained in subsection~\ref{sec:FFs--sensitive processes}.
Additionally, it is observed that the relative errors for the FFs of heavy quarks ($c$ and $b$) and gluon remain almost unchanged throughout the entire region depicted in the plot. This stability is expected, given that the data sets from the heavy-flavor tagged measurements of $e^+e^-\to c\bar{c}/b\bar{b}$ and $H\to c\bar{c}/b\bar{b}$ processes can lead to well-constrained FFs from heavy flavors.
Upon the removal of the data sets from the Higgs boson decays, it is evident that the measurements from the decay of the Higgs boson exert a profound influence on the FFs of gluon. This is because the data from the decays of the Higgs boson to gluons represent a unique channel for the fragmentation of gluon into three light charged hadrons. Furthermore, a slight increase in the relative error for heavy quark FFs in the small $x$ region is observed. This is a consequence of the exclusion of data from the $H\to c\bar{c}/b\bar{b}$ processes.

The data sets on quark pair production at lepton colliders show a strong influence on the FFs of light quarks, which has been explained in subsection~\ref{sec:FFs--sensitive processes}. Furthermore, it can be observed that the relative errors for the FFs to light charged hadrons from heavy quarks are slightly larger than those of the baseline fit, particularly in the small $x$ region.
This is analogous to the situation observed after subtracting data sets from Higgs boson decays.

\begin{figure}[t!]
	\centering
       \vspace{-2cm}
	\includegraphics[width=0.7\textwidth]{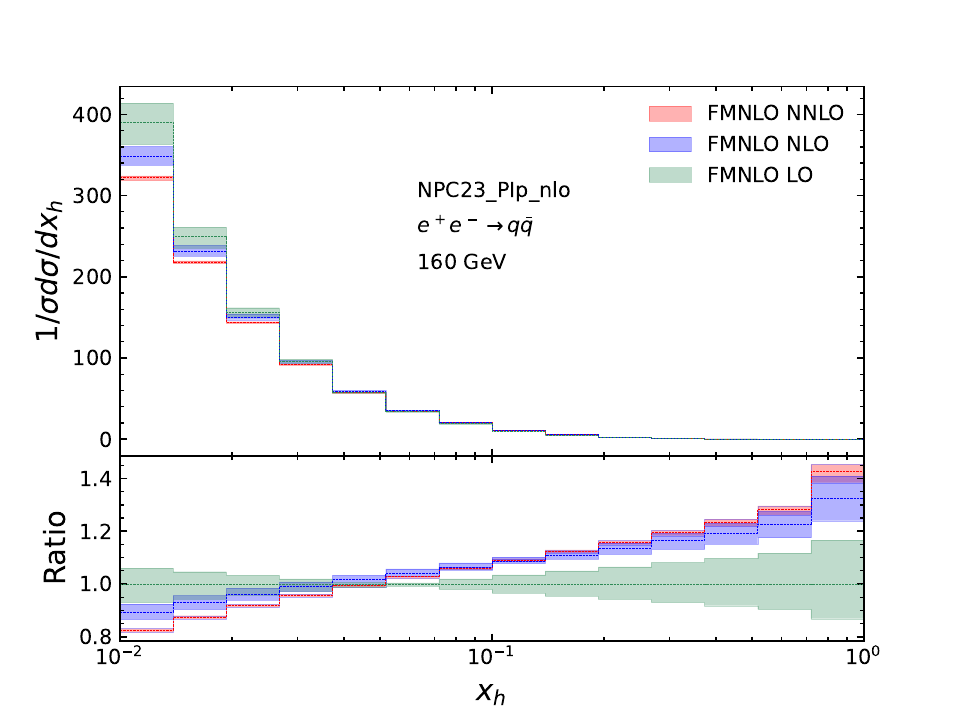}
    \includegraphics[width=0.7\textwidth]{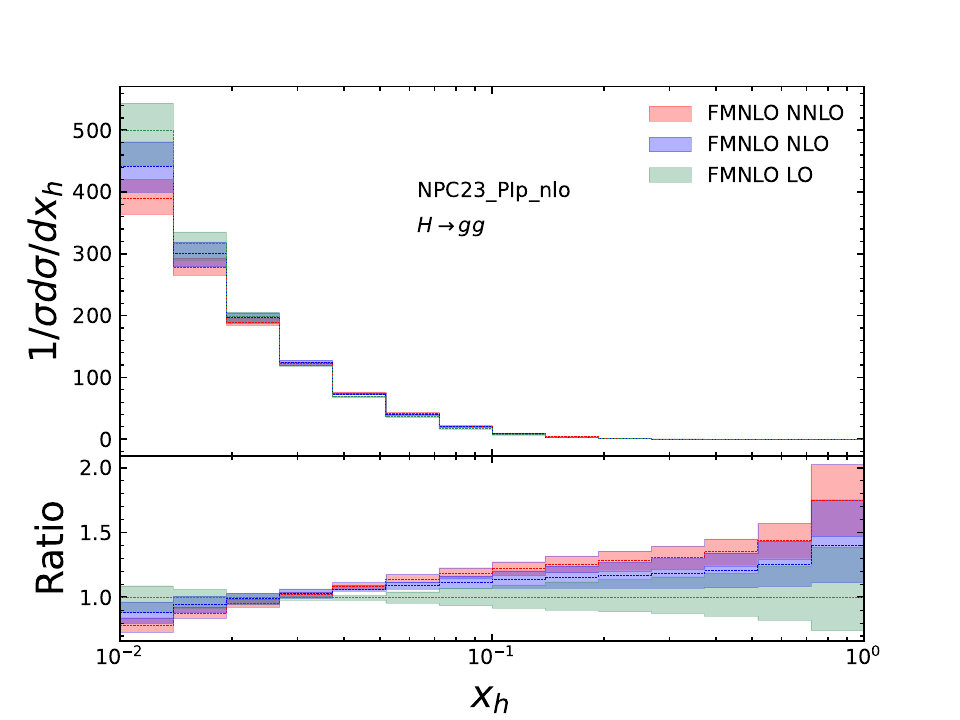}
	\vspace{-3ex}
	\caption{\label{figure:A4001_siatest_NPC23_PIp_nlo}The distribution of the hadron energy fraction from FMNLO at various orders for electron-positron annihilation at a center-of-mass energy of $160\,\GeV$, and for Higgs boson decaying into gluons. In the lower panel, the ratio to the central value at LO is presented. The error bands represent the scale uncertainties.}
\end{figure}

We now turn to study the effect of the higher-order QCD corrections on the FFs to light charged hardons.
We first show in Figure~\ref{figure:A4001_siatest_NPC23_PIp_nlo} the distribution of the hadron energy fraction for $q\bar{q}$ production at $160\,\GeV$, and for the Higgs boson decaying into gluons, where the partons subsequently undergo fragmentation to produce the identified hadron $\pi^+$. The FF set used for the evaluation of predictions at various orders is the NLO set fragmenting into $\pi^+$ from $\rm NPC23$. The NNLO corrections are calculated using FMNLO as explained in Appendix~\ref{Appendix:NNLO}.
The color bands represent QCD scale uncertainties.

\begin{figure}[h!]
	\centering
	\includegraphics[width=\textwidth]{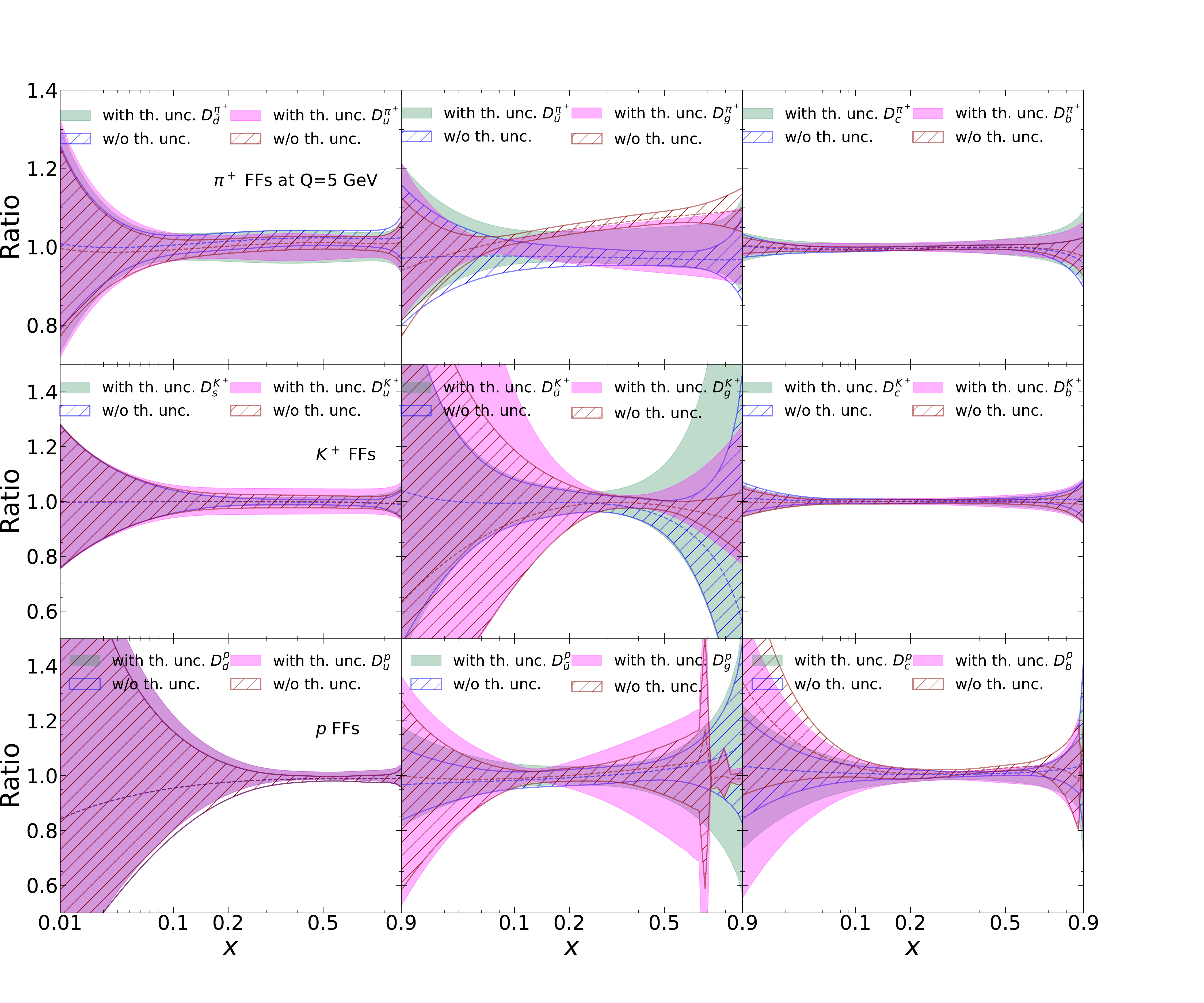}
	\vspace{-1.5cm}
	\caption{\label{figure:nounc_p_5GeV}
A comparison between the baseline fit, including the theoretical uncertainties, and a refitted version that excludes the theoretical uncertainties.
 }
\end{figure}

Figure~\ref{figure:A4001_siatest_NPC23_PIp_nlo} illustrates that the NLO prediction exhibits significantly reduced scale uncertainties in comparison to the LO prediction across the entire region of $x_h$.
The inclusion of NNLO corrections results in a further reduction of scale uncertainties. Furthermore, we also calculate the predictions using the NNLO set fragmenting into $\pi^+$ from $\rm NNFF1.0$~\cite{Bertone:2017tyb} as the input FFs and get similar results on the size of QCD corrections and scale uncertainties.
Consequently, we perform an alternative NLO fit by excluding the theoretical uncertainties to discuss the potential effects of the higher-order corrections.
All resulting FFs are compared to the baseline fit including the theoretical uncertainties and their ratios are shown in Figure~\ref{figure:nounc_p_5GeV}. The colored band represents the associated uncertainty estimated with the Hessian method at the 68\% confidence level.

\begin{figure}[t!]
	\centering
 \vspace{-.3cm}
	\includegraphics[width=0.9\textwidth]{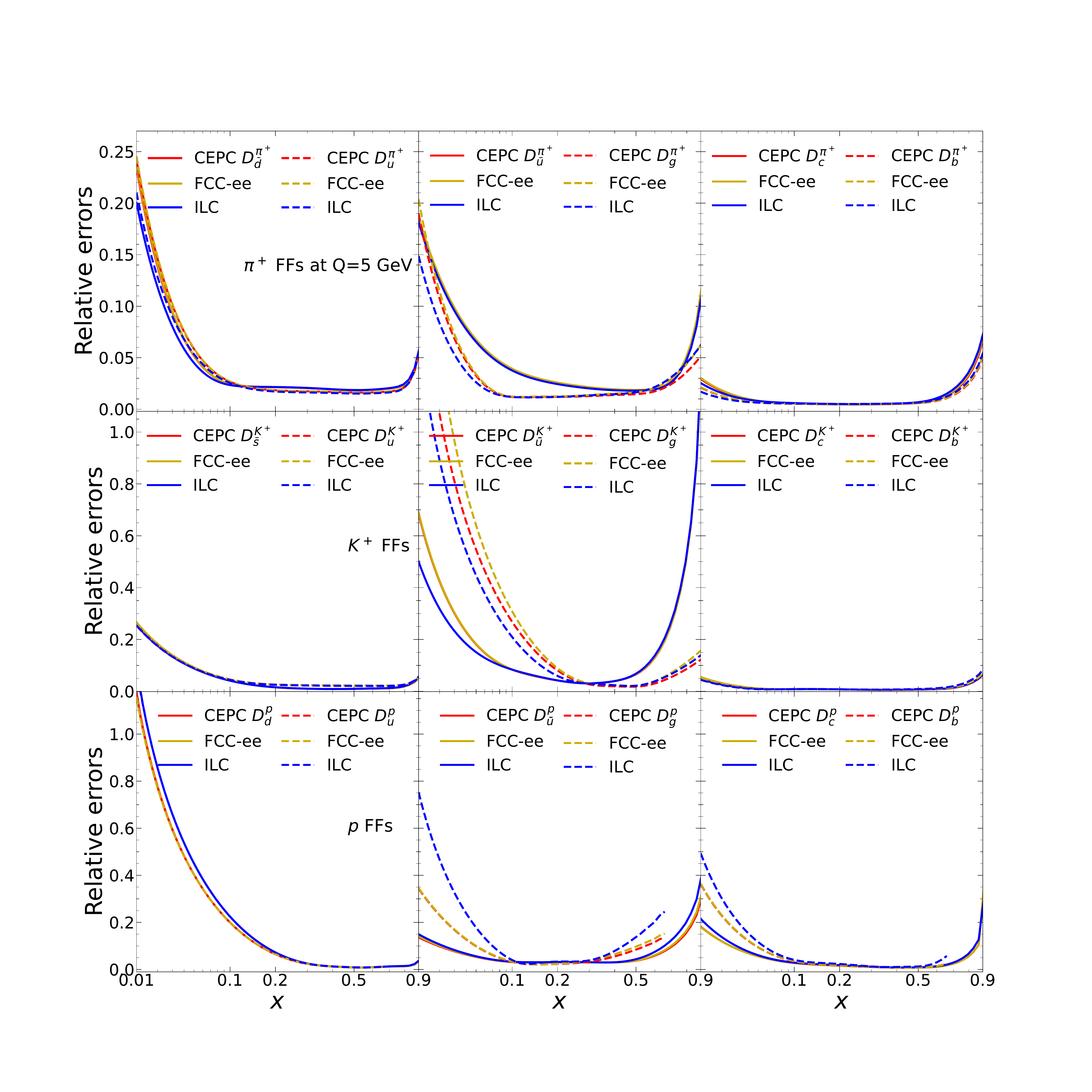}
	\vspace{-1.5cm}
	\caption{\label{figure:nounc_lepton_p_5GeV}
Similar to Figure~\ref{figure:lepton_p_5GeV}, but excluding theoretical uncertainties in fit of FFs.
 }
\end{figure}

As illustrated in Figure~\ref{figure:nounc_p_5GeV}, we find that excluding theoretical uncertainties has a significant impact and tends to reduce uncertainties of FFs, with the new error bands nearly within the Hessian uncertainties of the baseline fit.
For instance, the Hessian uncertainties for the $\bar{d}$ quark fragmenting into $\pi^+$ and the $u$ quark fragmenting into the $K^+$ are approximately two times smaller than those of the baseline fit. For the sub-figures in the second column, the error bands excluding theoretical uncertainties are considerably smaller than those of the baseline fit. In the case of heavy quarks, the error bands excluding theoretical uncertainties
show a significant reduction in uncertainty, as is evident from the FFs to the proton and $\pi^+$ in a wide kinematic range. Furthermore, Figure~\ref{figure:nounc_lepton_p_5GeV} presents a comparison of the relative errors among the FFs obtained from CEPC, FCC-$ee$, and ILC measurements without including theoretical uncertainties. Compared to the fits shown in Figure~\ref{figure:lepton_p_5GeV}, the corresponding results in Figure~\ref{figure:nounc_lepton_p_5GeV} are shifted downwards in the plot, indicating a stronger constraint on the FFs when higher-order corrections are included in the fit. Consequently, the evaluation of NNLO QCD corrections
for all processes shown in Table~\ref{tab:PseudoData} is necessary for future precision determination of FFs.

\subsection{The constraints from three-jet production}
The present study is primarily concerned with inclusive hadronic production at lepton colliders. Nevertheless, the experiment at a lepton collider is also capable of measuring hadron
multiplicity distributions in energy fraction for three-jet production. In three-jet events, where the jets are ordered according to their energies, the first two jets are enriched in quarks and the third in gluons. In this manner, the fragmentation of gluon can be investigated through the examination of the momentum spectrum of the lowest-energy jet (enriched in gluons) in three-jet events.

\begin{figure}[h!]
	\centering
	\begin{minipage}{0.49\linewidth}
		\centering
		\includegraphics[width=1.0\linewidth]{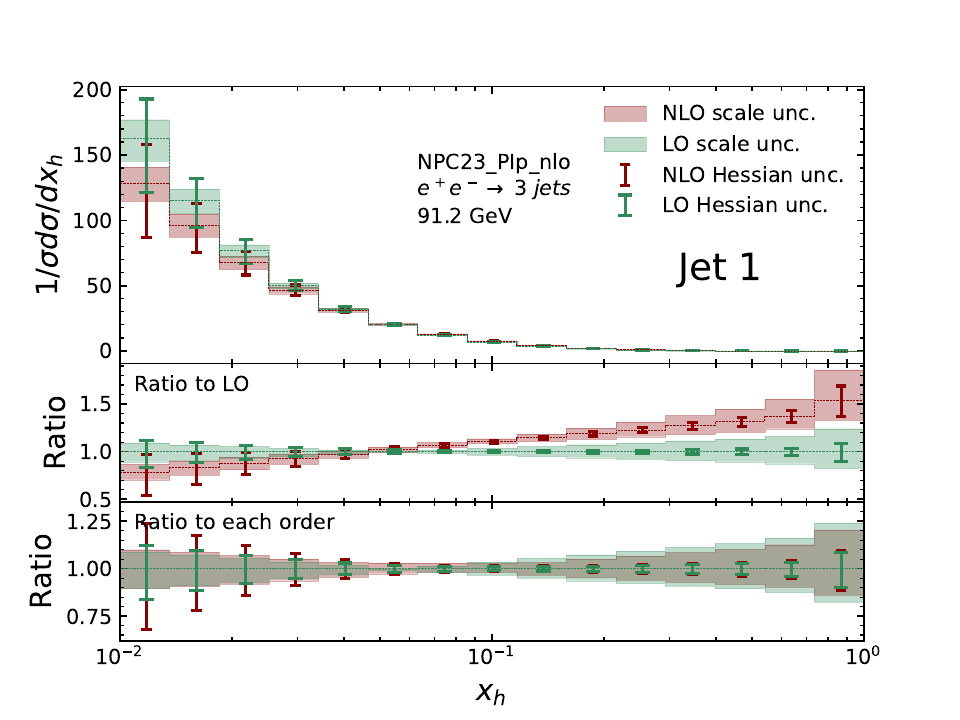}
\hspace{-1.5cm}
	\end{minipage}
	\begin{minipage}{0.49\linewidth}
		\centering
		\includegraphics[width=1.0\linewidth]{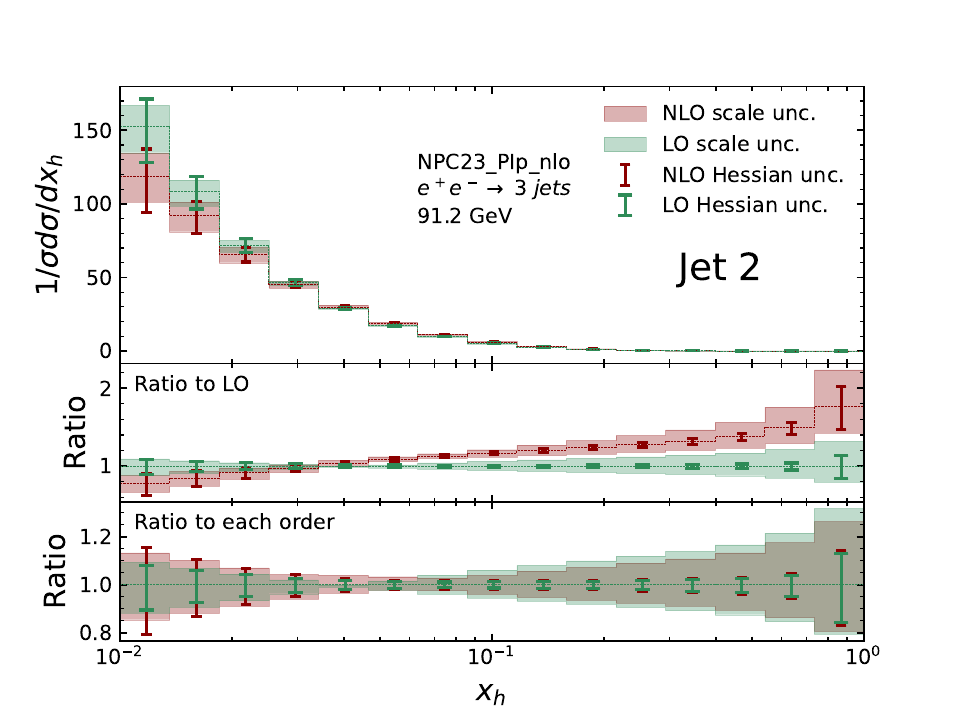}

	\end{minipage}

\vspace{-0.1cm}
	\begin{minipage}{0.49\linewidth}
		\centering
		\includegraphics[width=1.0\linewidth]{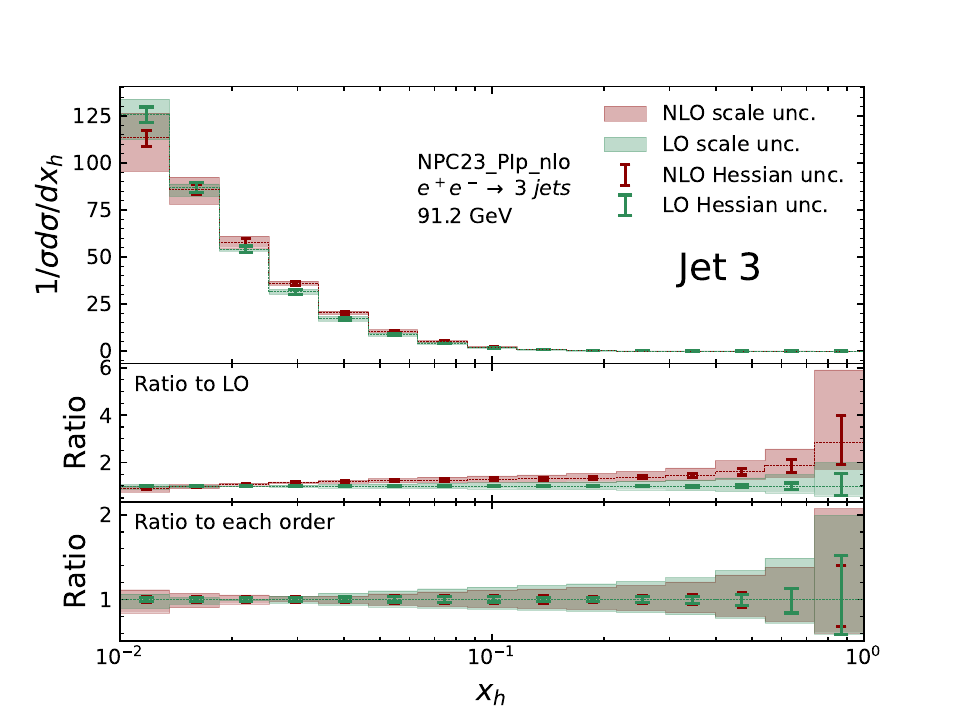}
	\end{minipage}
 \vspace{-0.4cm}

        \caption{A comparison between the scale uncertainties and the Hessian uncertainties for the distribution of the hadron energy fraction with a center-of-mass energy of $91.2\,\GeV$ in each jet of three-jet events. The jets were ordered according to their energies, with the upper-left, upper-right, and second-row plots, respectively. In the last two panels of each plot, the ratio to the central value at LO and the corresponding order are presented, respectively. The color bands represent scale uncertainties, while the error bars indicate Hessian uncertainties.}
 \label{figure:A5002_trijet}

\end{figure}

In a manner analogous to the approach employed by the ALEPH Collaboration in measuring the jets of three-jet events from hadronic $Z$ decays, the jets are clustered using the $k_\perp$ (Durham) algorithm with the E recombination scheme and a jet resolution parameter of $y_{cut} =0.01$. It is required that the polar angle between each jet and the beam axis be between $30^{\circ}$ and $150^{\circ}$.
Figure~\ref{figure:A5002_trijet} presents a comparison between scale and Hessian uncertainties for the distribution of the hadron energy fraction in each jet of three-jet events, ordered according to their energies. The fragmentation functions employed in this analysis are those derived from the NLO set fragmenting into $\pi^+$ from $\rm NPC23$. The color bands represent the scale uncertainties, while the error bars indicate the Hessian uncertainties. It can be observed that the NLO result for each jet of three-jet events exhibits a slightly reduced scale uncertainty compared to the LO result in the large $x_h$ region. However, it is notable that there is an unusual increase in scale uncertainty in the small $x_h$ region, as is evident from the final panel of each plot in Figure~\ref{figure:A5002_trijet}. This phenomenon can be attributed, at least in part, to the fact that the scale uncertainty at the LO level is influenced solely by the fragmentation scale, whereas the NLO level is additionally affected by the renormalization scale.

In the case of the first two jets, it can be demonstrated that the Hessian uncertainties are larger than the scale uncertainties in the small $x_h$ region. Moreover, as the value of $x_h$ increases, the scale error band gradually encompasses the Hessian error bar. In the case of the lowest-energy jet in three-jet events, the scale uncertainties exceed the Hessian uncertainties across the entire $x_h$ region depicted in the plot. This observation is evident in the second panel of each plot in Figure~\ref{figure:A5002_trijet}.
The ALEPH Collaboration has measured the inclusive cross section for various particles in three-jet events, and both systematic and statistical errors are relatively large. This is reported in Ref.~\cite{ALEPH:1999udi}. Nevertheless, it is anticipated that the corresponding error will be significantly reduced at future lepton colliders. Consequently, when a global fit including the data from three-jet events is performed to further enhance the understanding of the FFs of partons, particularly those of gluon fragmenting into $\pi^{\pm}, K^{\pm}$ and $p/\bar{p}$, it is necessary to calculate higher-order corrections for three-jet production at electron-positron colliders.

\section{Summary}
\label{sec:summary}
In this work, we have studied the constraints that future lepton colliders will impose on the FFs to light charged hadrons from quarks and gluon in the framework of QCD collinear factorization.
We find the data from the CEPC has the potential to significantly reduce the uncertainties of FFs in a wide kinematic region, especially for quark FFs. For example, the relative errors of FFs for the constituent quark $\bar{d}$ fragmenting to $\pi^+$ are reduced by almost a factor 5, from around 20\% to a few percent, in a wide range of momentum fractions, compared to those from the NPC23 global analyses based on current world data. For the gluon FF, we observe improvements of uncertainties to all charged hadrons, except for gluon fragmenting to $\pi^+$ in the small $x$ region due to our selection cuts. We also study the cases of the FCC-$ee$ and the ILC, and the conclusions on the reduction of FF uncertainties are similar to those of the CEPC.

We have conducted alternative fits by systematically excluding a specific group of data sets at a time and re-fitting the FFs, to assess the influence of specific data sets on distinct fragmentation processes. Taking the fit for the CEPC as an example, we find the measurements of $W^-W^{+*}$ production impose strong constraints to the FFs of light (anti-)quarks while having no effects on gluon FFs. In the case of removal of the Higgs boson decay processes, we find they show a large impact on the FFs of gluon, which is not so surprising since the data from the $H\to gg$ process is a unique channel to constraining gluon fragmentation to three light charged hadrons among the full data sets listed in Table~\ref{tab:PseudoData}.
The $q\bar{q}$ production at various center-of-mass energy of course provides the dominant constraints in general.

In addition, we conduct NLO fits without including the theoretical uncertainties to study the impact of higher-order QCD corrections. We observe a significant reduction of the FFs error bands in all cases when excluding theoretical uncertainties.
For example, the Hessian uncertainties for $\bar{d}$ quark fragmenting into $\pi^+$ and $u$ quark fragmenting into $K^+$ are about 2 times smaller than those of the baseline fit including the theoretical uncertainties.
Therefore, the evaluation of NNLO QCD corrections for all processes shown in Table~\ref{tab:PseudoData}
are necessary for future precision determination of FFs. Currently, the NNLO QCD corrections are available for both SIA and the decays of the Higgs boson to gluons, and have been implemented into the FMNLO program as described in the appendix.

Finally, we discuss the impact of measurements of three-jet production on the FFs, which may further improve the constraints on gluon FFs. By comparing the scale variations of theoretical predictions and uncertainties induced by FFs, we conclude that the NNLO corrections will be needed in the future to effectively use the data for analyses of FFs.

\subsection*{Acknowledgements}
We are grateful to ChongYang Liu for providing the drawing script.  The work of JG is supported by the National Natural Science Foundation of China (NSFC) under Grant No.~12275173.

\appendix

\begin{figure}[t!]
	\centering
        \vspace{-.3cm}
	\includegraphics[width=\textwidth]{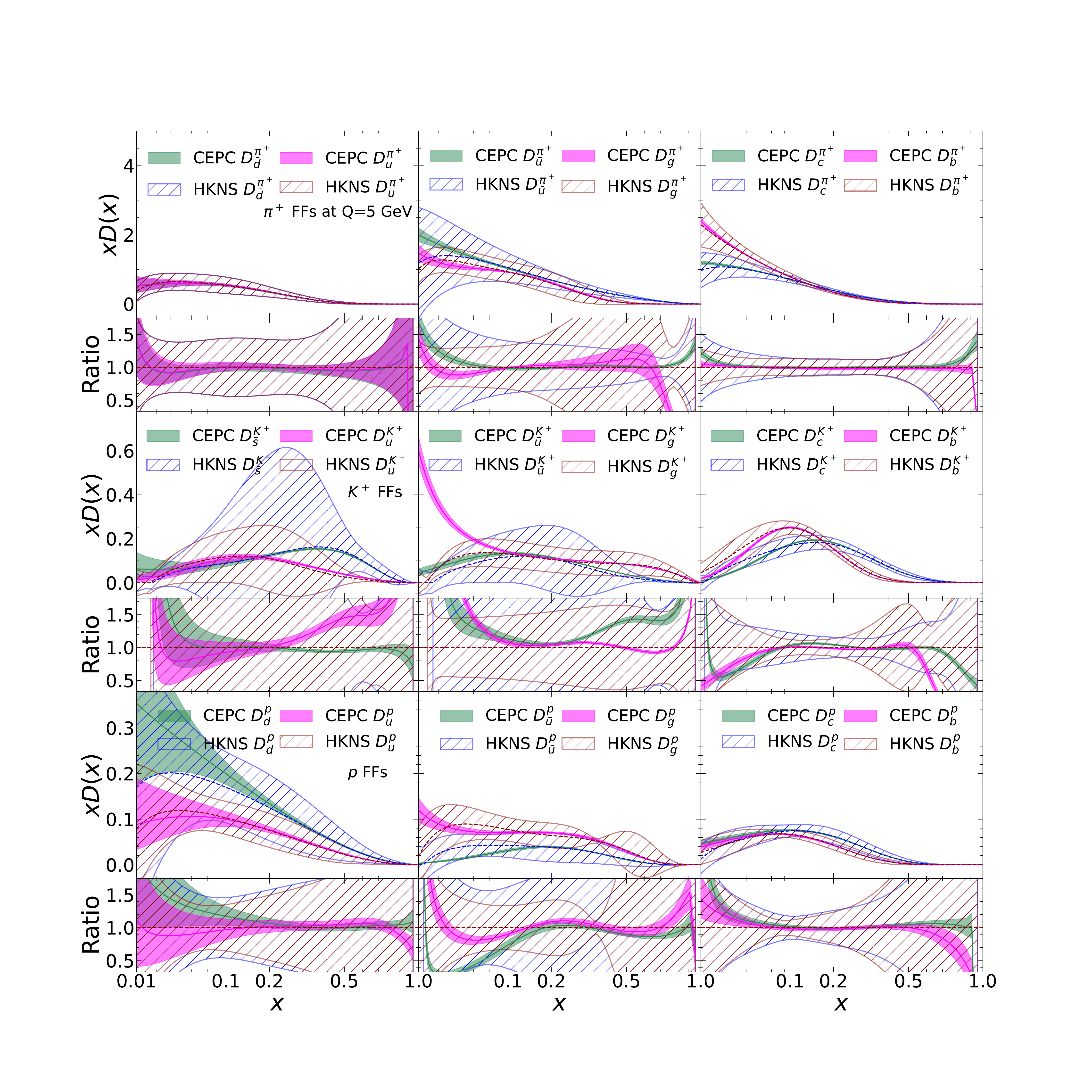}
	\vspace{-1.5cm}
	\caption{\label{figure:comHKNS_asym_p_5GeV}
A comparison of our fragmentation functions to those of HKNS on the diverse outcomes of parton fragmentation, including $\pi^+$, $K^+$, and $p$, at an energy of $5 \,\GeV$. The colored band represents Hessian uncertainties, with their ratios normalized to the central value of the baseline from the HKNS study displayed in the lower panel of each subplot.}
\end{figure}

\section{The analysis based on HKNS sets}
\label{Appendix:HKNS}
In this Appendix,
the pseudo--data enumerated in Table~\ref{tab:PseudoData} are generated using $\rm HKNS$ sets~\cite{Hirai:2007cx}, which are extracted from SIA data only.
We utilize the FFs from the CEPC as a case study to investigate the extent to which the pseudo--data constrain the FFs.
A comparison of the baseline $\rm HKNS$ FFs with those based on CEPC pseudo--data at the scale $Q_0 = 5 \,\GeV$ is presented in Figure~\ref{figure:comHKNS_asym_p_5GeV}. For simplicity, only the FFs for various partons fragmenting into $\pi^+, K^+$, and $p$ are shown. The FFs for negative charge hadrons are obtained through charge conjugation. The colored band represents Hessian uncertainties, with their ratios normalized to the central value of the baseline from the HKNS study displayed in the lower panel of each subplot.

As anticipated, the central values of $\rm HKNS$ and our fit exhibit a good agreement, spanning a broad range of $x$ values in all cases.
However, compared to Figure~\ref{figure:comNPC23_asym_p_5GeV}, there are more significant deviations in the central values, such as the FF to $K^+$ from gluon at $x<0.1$. This can be attributed to the fact that the HKNS parameterization form of fragmentation functions to charged hadrons is different from the form used in our fit framework.
For the quarks fragmenting into light charged hadrons, we observe a marked reduction in the FFs uncertainties. This is of significance for the $c$ and $b$ quarks, for the reason that a considerable number of measurements from heavy-flavor tagged hadronic events in $e^+e^-$ collisions have strong constraints to heavy quark FFs.
In the cases of the constituent quarks and sea quarks, there is a reduction in the FFs uncertainties across a wide range of $x$. This is of particular significance for the FFs to $K^+$ from the $u$ and $\bar{s}$ quarks.
The primary reason for this improvement is the increased statistics and the diverse measurements across a wide range of collision energies, from the $Z$-pole to $360\,\GeV$, which will be available at the CEPC.

In the case of the gluon, it can be observed that the CEPC can significantly reduce the uncertainties of FFs, compared to the $\rm HKNS$ results.
It is also observed that the error bands from $\rm HKNS$ sets encompass those from our fit, spanning a broad range of $x$ values in all cases.
These results highlight the importance of the measurements for the $H\to gg$ channel which represents a unique handle on the poorly known gluon FF to three light charged hadrons in our work.
Furthermore, it can be seen that the two bands for the FFs to $K^+$ from the gluon even do not overlap below $x<0.05$. This can be attributed to the insufficient number of data points available for the determination of gluon FFs to $K^+$ after selections in the small $x$ region.

In conclusion, the data from the CEPC has the potential to significantly reduce the uncertainties of FFs in a wide kinematic region, regardless of whether we generate pseudo--data using the $\rm HKNS$ sets or the $\rm NPC23$ sets extracted from a global analysis, using SIA, SIDIS, and pp data.

\section{NNLO calculations in FMNLO}
\label{Appendix:NNLO}

Through this study we have developed \texttt{v2.1} of the \texttt{FMNLO} program, which allows for NNLO calculations and grid generations for hadron multiplicities in SIDIS, SIA, and decays of the Higgs boson to gluons, that can be found on the website\footnote{\url{http://fmnlo.sjtu.edu.cn/~fmnlo/}}.
The implementation is based on the analytical results from original calculations in Refs.~\cite{Goyal:2023xfi,Bonino:2024qbh,Rijken:1996vr,Rijken:1996npa,Rijken:1996ns,Mitov:2006wy,Soar:2009yh,Almasy:2011eq}.
Instructions on installation and usage of \texttt{FMNLO} can be found in Appendix A of Ref.~\cite{Liu:2023fsq}.
Here we highlight only the usage of the NNLO component, which has been available since \texttt{v2.1}.
%
We take the module used for the SIA calculation at NNLO mentioned in subsection~\ref{sec:Alternative fit} as an example.
This module, named \texttt{A4001}, is one of the default examples available in the \texttt{FMNLOv2.1} package.
The parameter card for this module corresponds to the file \texttt{FMNLOv2.1/mgen/A4001/proc.run}, and reads
\begin{verbatim}
sidis A4001
# subgrids with name tags
grid siannlo240
pdfname	'CT14nlo'
etag 'e+'
htag 'p'
obs 3
zdef 2
cut 0.02
q2d 10.0
q2u 100000.0
xbjd 0.14
xbju 0.18
yid 0.3
yiu 0.5
pdfmember 0
sqrtS 240.0
Rscale 1.0
Fscale 1.0
ncores 30
maxeval	1000000
iseed 11
end
\end{verbatim}
where
\begin{itemize}
\item \texttt{sidis} specifies the name of the directory that contains the module to be loaded.
\item \texttt{grid} is a string indicating the name of the running job.
\item \texttt{obs} specifies different observables to be calculated: \texttt{3, 4} is for NNLO calculation of multiplicity distribution in hadron energy fraction for SIA and decays of the Higgs boson to gluons, respectively. For these two cases, parameters from \texttt{pdfname} to \texttt{htag}, and from \texttt{zdef} to \texttt{pdfmember} are irrelevant; \texttt{2} is for NNLO calculation of multiplicity distribution in hadron scaled momentum ($z$) for SIDIS. Here \texttt{zdef} is irrelevant while other input parameters follow descriptions in the appendix of Ref.~\cite{Gao:2024dbv}.
\item \texttt{sqrtS}: the c.m. energy $\sqrt{s}$ or the mass of the Higgs boson in GeV.
\item \texttt{Rscale}, \texttt{Fscale}: ratios of the renormalization and factorization scale w.r.t. our nominal choice of $Q$. Again \texttt{Fscale} is only relevant for SIDIS calculation with \texttt{obs=2}.
\item \texttt{ncores, maxeval, iseed} indicate technical parameters of numerical calculations, including number of CPU cores used, maximum number of integrand evaluations, and seed for pseudo-random-number generation.
\end{itemize}
After the grid generation, the multiplicity distributions can be obtained for arbitrary FFs and binnings following the same procedures as described in Ref.~\cite{Liu:2023fsq} for \texttt{FMNLOv1.0}. We calculate the K-factors up to NNLO for SIA and SIDIS and compare them with the results obtained from private codes in Ref.~\cite{Li:2024etc} and the left plot in Figure 2 shown in Ref.~\cite{Bonino:2024qbh}. We find good agreement between them.

\section{Three-jet production in FMNLO}
\label{sec:fmnlo}

In \texttt{FMNLOv2.1} we also include a new module \texttt{A5002} for NLO calculation of hadron multiplicity distributions in energy fraction for three-jet production at electron-positron collisions.
Note in the calculation the nominal choice of fragmentation scale is set to the energy of individual jet while the choice of renormalization scale follows the options in \texttt{MG5\_aMC@NLO} \cite{Alwall:2014hca,Frederix:2018nkq}.
This module is tailored for the measurement on three-jet production from ALEPH collaboration~\cite{ALEPH:1999udi} as described in the main text, or from L3 collaboration~\cite{L3:1995rnv}.
Though a general case with different jet algorithms and selection cuts can be implemented easily.
The parameter card for this module corresponds to the file \texttt{FMNLOv2.1/mgen/A5002/proc.run}, and reads
\begin{verbatim}
process A5002
# subgrids with name tags
grid trijet1_ag1_e91
obs 1
cut 0.02
ptj1 0.0
ptj2 100.0
# 1/2 for kT or LUCLUS
jalg 1
# 1/2/3 for individual jets ordered in E
ifor 1
# in MG5 format
set ptj 1.0
set lpp1 0
set lpp2 0
set ebeam1 45.59
set ebeam2 45.59
set iseed 11
set muR_over_ref 1.0
set muF_over_ref 1.0
set fixed_ren_scale  True
set muR_ref_fixed 91.18
set req_acc_FO 0.002
end

\end{verbatim}
The additional parameters compared to similar modules presented in \texttt{FMNLOv1.0} are
\begin{itemize}
\item \texttt{jalg} specifies the jet algorithm and selection cuts used:
currently \texttt{1, 2} is tailored for the ALEPH and L3 measurements respectively.
\item \texttt{ifor} indicates the jet to be analysed: can be  \texttt{1, 2, 3} corresponding to the \texttt{1-st, 2-nd, 3-rd} jet ordered according to jet energies, namely \texttt{ifor = 1} corresponds to the leading jet in energies.
%
\end{itemize}

\bibliographystyle{JHEP}
\bibliography{cite}

\end{document}